\documentclass[review]{elsarticle}
\usepackage[lofdepth,lotdepth]{subfig}
\usepackage{lineno,hyperref,tabularx}
\usepackage{graphicx,color,amsmath,amsthm,amssymb,bm,array}
\graphicspath{{./}}
\newcommand\Gr{\mbox{\textit{G}}}
\newcommand{\Rey}{\mbox{\textit{Re}}}
\modulolinenumbers[5]

\makeatother

\journal{International Journal of Heat and Mass Transfer}

\begin{document}

\begin{frontmatter}

\title{Conductive and convective heat transfer in fluid flows between differentially heated and rotating cylinders}


\author[mymainaddress,mysecondaryaddress]{Jose M. Lopez\corref{mycorrespondingauthor}}
\ead{jose.lopez@fau.de}

\author[mymainaddress]{Francisco Marques}

\author[mysecondaryaddress]{Marc Avila}
\cortext[mycorrespondingauthor]{Corresponding author}

\address[mymainaddress]{Department of F\'{i}sica
  Aplicada, Universitat Polit\`{e}cnica de Catalunya, Girona s/n,
  M\`{o}dul B4 Campus Nord, 08034 Barcelona, Spain}
\address[mysecondaryaddress]{Institute of Fluid Mechanics,
  Department of Chemical and Biological Engineering,
  Friedrich-Alexander-Universit\"at Erlangen-N\"urnberg, 91058
  Erlangen, Germany}

\begin{abstract}

The flow of fluid confined between a heated rotating cylinder and a 
cooled stationary cylinder is a canonical experiment for the study of 
heat transfer in engineering. The theoretical treatment of this system 
is greatly simplified if the cylinders are assumed to be of infinite
length or periodic in the axial direction. In these cases heat
transfer in the laminar regime occurs only through conduction as in a solid. 
We here investigate numerically heat transfer and the onset of
turbulence in such flows by using both periodic and no-slip boundary 
conditions in the axial direction. 
The influence of the geometric parameters is 
comprehensively studied by varying the radius ratio ($0.1 \leq \eta
\leq 0.99$) and the length-to-gap aspect ratio ($5 \leq \Gamma \leq
80$). Similarly, a wide range of Prandtl, Rayleigh, and Reynolds
numbers is explored ($0.01 \leq \sigma \leq 100$, $Ra \leq 30000$, and
$Re \leq 1000$, respectively). We obtain a simple criterion,
$Ra\lesssim a(\eta)\Gamma$, which determines whether the
infinite-cylinder assumption can be employed. The coefficient $a$ is
well approximated by a cubic fit over the whole $\eta$-range. 
Noteworthy the criterion is independent of the Prandtl number 
and appears robust with respect to Reynolds number even  
beyond the laminar regime.

\end{abstract}


\end{frontmatter}

\nolinenumbers

\section{Introduction}

Instabilities driven by the combination of rotation and thermal
gradients determine the dynamics of complex geophysical, astrophysical
and industrial flows. Simple models of such flows can be tested in
laboratory experiments of a laterally heated differentially rotating
annulus, for which extensive information about the physical mechanisms 
and flow regimes can be found in the literature (see~\cite{La12} for a
review on this topic). The case of rotating heated inner cylinder and 
stationary cooled outer cylinder (RHISCO) is a model for the cooling
of rotating machinery, the  solidification of pure metals, techniques
of chemical vapor deposition, rotating-tube heat exchangers and
nuclear reactor fuel rods~\citep{kreith1968,Singer1984,vives1988,SelOz14}.  
The geometry of such an experimental apparatus is fully specified by 
the length-to-gap aspect ratio $\Gamma=h/(r_o-r_i)$, 
and the radius-ratio  $\eta=r_i/r_o$, 
where  $r_i$ and $r_o$ are the radii of the inner and outer cylinders, 
and $h$ is their height. 

Depending on the geometry, RHISCO experiments used in the literature can be classified in two groups. The first group of experiments~\citep{SK64,SoCo79,LeGoPriMu08} is characterized by long cylinders $\Gamma=h/(r_o-r_i)\ge 100$ and narrow gap $\eta=r_i/r_o\lesssim 1$. Ali and Weidman~\cite{AlWe90} performed a detailed linear stability analysis of such flows using axial periodicity and reported on the influence of the Prandtl number ($\sigma$) and $\eta$ on the stability boundaries. Their results showed a good agreement with~\cite{SK64} and, to a lesser extent with~\cite{SoCo79}. Ali and Weidman attributed the discrepancies to the limitations of linear stability theory and the infinite-cylinder idealisation to capture the experimental details. A similar linear stability analysis~\citep{YoNaMu13} reported good agreement between numerical and related experimental results~\citep{LeGoPriMu08}. Nonlinear simulations for small temperature gradients were provided by Kedia~\emph{et al}.~\cite{KeHuCo98} who quantified the heat transfer across the system. A second group of experiments embraces apparatuses with moderate aspect ratio and wide gap. Ball and Farouk~\cite{BaFa87,BaFa88,BaFa89} reported heat transfer measurements as well as the sequence of flow transitions using an experimental setup with $\Gamma=31.5$ and $\eta\sim 0.5$. Subsequent numerical simulations~\citep{KuBa97} for $\Gamma=10$ and $\eta=0.5$ provided insight on the bifurcations structure of the system. However, the results showed significant
discrepancies with experiments suggesting strong effects of the axial boundaries.

An accurate numerical simulation of the axial (Ekman) boundary layers in flows between long cylinders entails a substantial computational cost, especially for rapid rotation and large temperature gradients. The assumption of axial periodicity reduces the computational effort because the Ekman layers are not present and only a short central fraction of the apparatus needs to be simulated to estimate transport properties. Moreover, under this assumption variables can be expanded as a Fourier series in the axial direction, which greatly simplifies the numerical approach and enables the use of more efficient solvers. 

In this paper we determine under what conditions periodic boundary conditions can be employed to describe the dynamics of RHISCO experiments. In particular, we provide criteria to distinguish flow features that arise from the interplay between differential rotation and temperature, from those which are mainly determined by the axial boundaries or end walls. We compare the flow dynamics by using both physical (no-slip) and axially periodic boundary conditions in our numerical simulations. We show that axial periodicity renders a good approximation of laboratory flows as long as the Rayleigh number $Ra$ is small. We provide a simple criterion that determines whether heat transfer in the laminar flow is conductive or convective. In particular, conductive profiles, which enable the use of axially periodic boundary conditions, are obtained as long as $Ra< a(\eta)\Gamma + b(\eta)$. In addition, the coefficient $b$ may be neglected for $\Gamma\gtrsim15$.

\section{Specification of the system and numerical methods}\label{sec:spec}

\begin{figure}
  \begin{center}
    \begin{tabular}{cc}
      $(a)$ & $(b)$\\
    \includegraphics[width=0.48\linewidth]{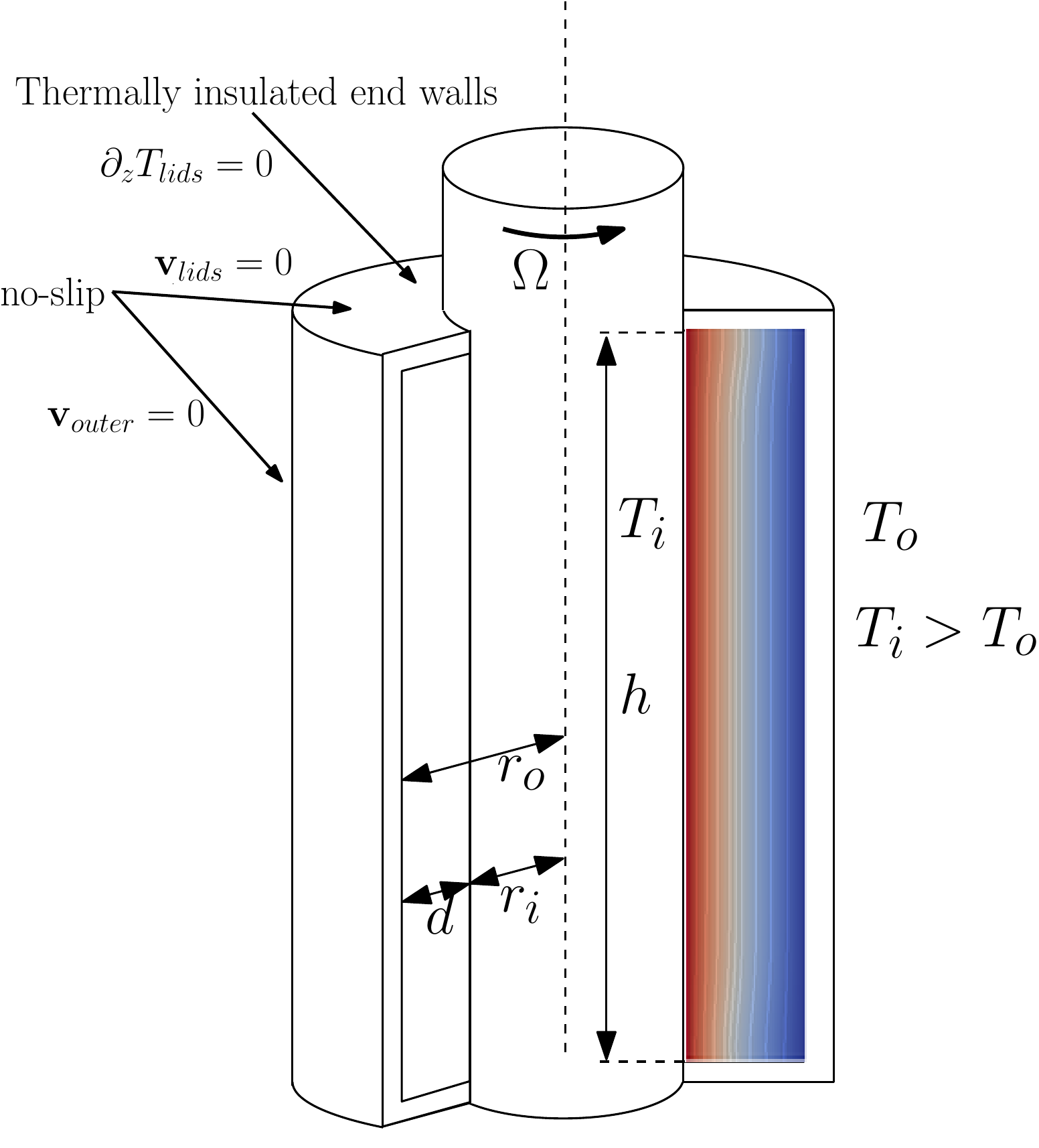}&
    \includegraphics[width=0.48\linewidth]{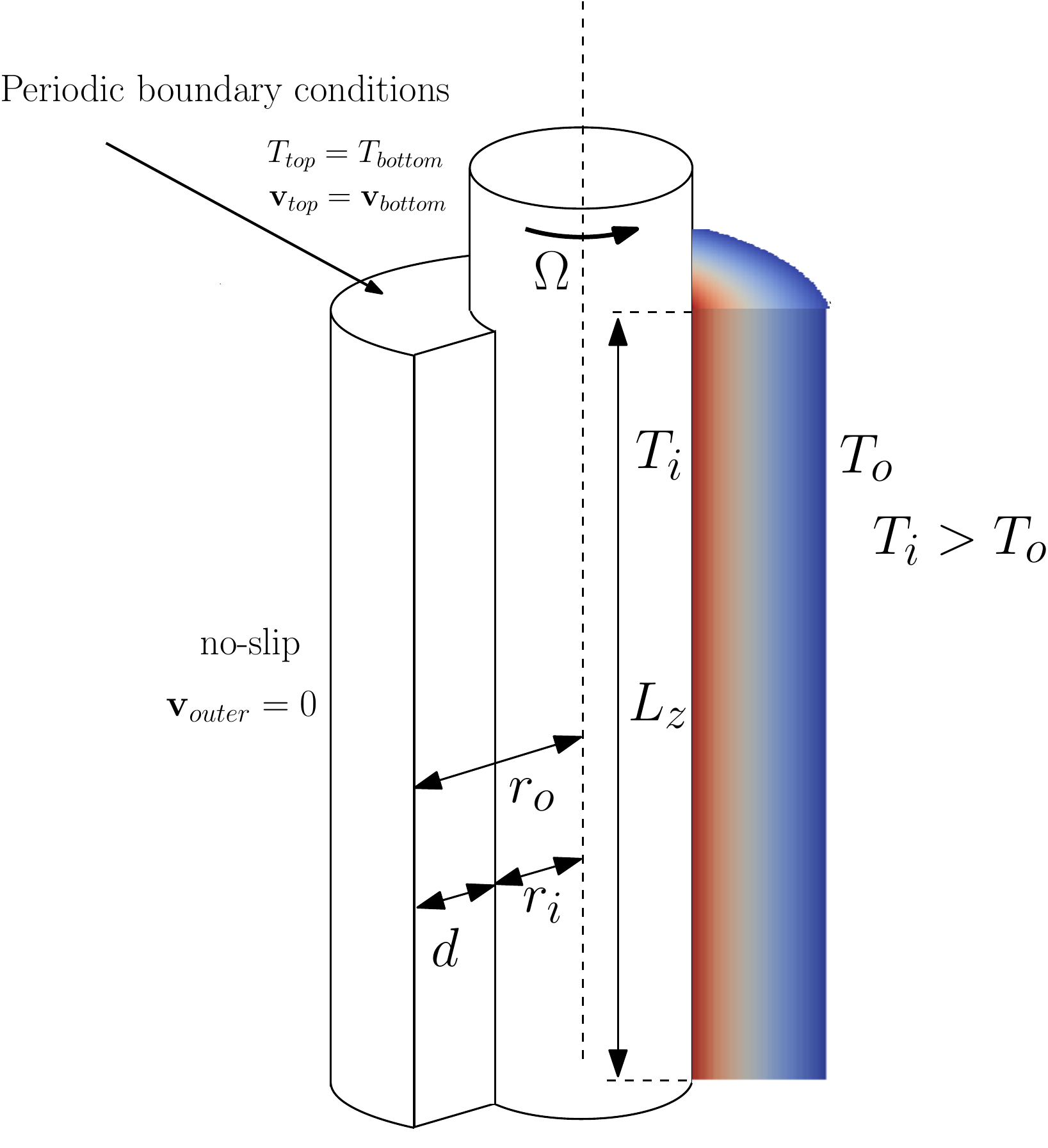}
    \end{tabular}
    \end{center}
\caption{ (color online) Sketches of geometry and boundary conditions
  for the two cases considered in this paper.  $(a)$ Stationary
  insulating endwalls (no-slip boundary condition for the velocity and
  zero flux for the temperature) and $(b)$ axially periodic boundary
  conditions. The temperature profile is superimposed on the right
  hand side of each figure.}
  \label{schematic}
\end{figure}

We consider the motion of an incompressible fluid of kinematic
viscosity $\nu$  and  thermal diffusivity $\kappa$ confined in the
annular gap between two rigid and concentric rotating cylinders of
radii $r_i$ and $r_o$. The inner cylinder rotates at constant angular
velocity $\Omega$, whereas the outer cylinder is kept at rest. A
radial thermal gradient is considered by setting the inner and outer
cylinder temperature to $T_i=\overline{T}+\Delta T /2$ and
$T_o=\overline{T}-\Delta T/2$, respectively. Here $\overline{T}$ is
the mean temperature of the fluid. The axis of the cylinders is
vertical, i.~e.~parallel to the gravitational acceleration $g$. We
study flows with stationary end walls and with axially periodic
boundary conditions (Figures~\ref{schematic} $(a)$ and $(b)$
respectively). 
The latter model the case of infinitely long
cylinders, whereas the former reproduce experimental boundary
conditions (no-slip for the velocity and thermally insulating end
walls for the temperature).  

\subsection{Governing equations} 

We consider the Boussinesq approximation including centrifugal buoyancy  in an inertial reference frame as described in~\cite{LoMaAv13}. The dimensionless governing equations are  
\begin{subequations}\label{formInon}
\begin{align}
  & (\partial_t+\mathbf{v}\cdot\nabla)\mathbf{v}=-\nabla p
    +\nabla^2\mathbf{v}+G T{\bf\hat z}
    +\epsilon T\mathbf{v}\cdot\nabla\mathbf{v},\label{formI_3v}\\
  & (\partial_t+\mathbf{v}\cdot\nabla)T=\sigma^{-1}\nabla^2T,\label{formI_3T}\\
  & \nabla\cdot\mathbf{v}=0,\label{formI_3inc}
\end{align}
\end{subequations}
where $\mathbf{v}=(u,v,w)$ denotes the velocity field vector and $T$ is the deviation of the temperature with respect to $\overline{T}$. The length, time, temperature and pressure scales chosen to make the set of equations dimensionless are the gap width $d=r_o-r_i$, the viscous time $d^2/\nu$, the temperature difference between the cylinders  $\Delta T$ and $(\nu/d)^2$ respectively. There are six independent dimensionless numbers (see table~\ref{nondim_numbers}). The term $\epsilon T\mathbf{v}\cdot\nabla\mathbf{v}$ accounts for centrifugal buoyancy, including secondary effects stemming from differential rotation or strong internal vorticity~\cite{LoMaAv13}. The equations are solved in cylindrical coordinates $(r,\theta,z)$

\begin{table}
  \begin{center}
    \begin{tabular}{|c|c|c|c|}
      \hline
      \text{Name} & \text{Symbol} & \text{Formula} & \text{Operation range} \\\hline
      \text{Grashof number} & $G$ & $\alpha g\Delta Td^3/\nu^2$ & $0 \le 
      G \le 12000$\\\hline
      \text{Relative density variation} & $\epsilon$ & $\alpha
      \Delta T =\Delta \rho / \rho_0$ &  $0 \le \epsilon \le 0.025$\\\hline
      \text{Prandtl number} & $\sigma$ & $\nu/\kappa$ &  $0.01 \le 
      \sigma \le 100$\\\hline
      \text{Aspect ratio} & $\Gamma$ & $h/d$ &  $5 \le 
      \Gamma \le 80$\\\hline
      \text{Radius ratio} & $\eta$ & $r_i/r_o$ &  $0.1\le \eta \le0.99$\\\hline
      \text{Reynolds number} & $\Rey$ &
      $\Omega r_i d/\nu$
      &  $0 \le \Rey \le 1000$\\\hline
      
    \end{tabular}
  \end{center}
  \caption{Dimensionless parameters. Here $\kappa$ and 
    $\alpha$ are respectively the thermal                                                                                                                                                                                                                                    
    diffusivity and the coefficient of volume expansion of the
    fluid, $\Delta\rho$ is the density variation associated with a
    temperature change of $\Delta T$. Last column indicates the range of values covered by
    the simulations shown in this paper.}
  
  \label{nondim_numbers}
\end{table}

\subsection{Numerical methods}

In the axially periodic case the onset of instabilities was determined via linear stability analysis of the basic flow as in~\cite{LoMaAv13}. Fully nonlinear simulations were performed using the Boussinesq-approximation~\cite{LoMaAv13}, which was added with the heat equation to the finite-difference-Fourier--Galerkin (hybrid MPI-OpenMP) code of Shi \emph{et al}.\cite{LiRaHoAv14}. A time-step $\delta t=2\times 10^{-5}$ viscous time units was used in all computations.    

For rigid end walls the governing equations were solved using a second-order time-splitting method. A pseudo-spectral formulation is used for the spatial discretisation, with the Fourier--Galerkin method in the azimuthal coordinate $\theta$ and Chebyshev collocation in $r$ and $z$. The code is based on a previous hydrodynamic code \cite{AGLM08}, which has been extended with the Boussinesq-approximation of~\cite{LoMaAv13} and parallelised as in \cite{LiRaHoAv14}. Details can be found in J.~M.~Lopez's PhD thesis~\cite{Lop15}. The numerical resolution has been chosen to ensure that the infinite norm of the spectral coefficients decays at least four orders in magnitude. Time steps as small as $\delta t=1\times 10^{-5}$ viscous time units have been required for numerical stability and accuracy of the second-order temporal scheme.                                               

\section{Conductive and convective basic flows}\label{sec:bf}

The assumption of axial periodicity allows to considerably simplify the calculation of the basic flow. The radial velocity is zero and the rest of variables only depend on the radial component. Under these conditions an analytical solution can be found by imposing a zero axial mass flux. This reads
\begin{subequations}\label{basicflow}
  \begin{align}
    & v_b(r)=Ar+\frac{B}{r}, \label{ut} \\
    & w_b(r)=G\biggr(C(r^2-r_i^2)+\Big(C(r_o^2-r_i^2)+\frac{1}{4}(r_o^2-r^2)
      \Big)\frac{\ln(r/r_i)}{\ln\eta}\biggr), \label{uz} \\ 
    & T_b(r)= \frac{\ln(r/r_o)}{\ln\eta}-\frac{1}{2}, \label{tem} 
 \end{align}
\end{subequations}

The azimuthal velocity $v_b$ is the classical Couette flow, $T_b$ corresponds to the temperature in a conductive regime and $w_b$ is the axial velocity profile induced by gravitational buoyancy. The parameters $A$, $B$ and $C$ are
\begin{align}
  & A=\frac{\Rey_o-\eta \Rey}{1+\eta}, \qquad
    B=\eta \frac{\Rey-\eta \Rey_o}{(1-\eta)(1-\eta^2)}, \label{param_rot} \\
  & C=-\frac{4\ln\eta+(1-\eta^2)(3-\eta^2)}
    {16(1-\eta^2)\big((1+\eta^2)\ln\eta+1-\eta^2\big)},
\end{align}
and the non-dimensional radii are $r_i=\eta/(1-\eta)$, $r_o=1/(1-\eta)$. 
 
\begin{figure}
  \begin{center}
    \begin{tabular}{cc}
      $(a)$ & $(b)$\\
    \includegraphics[width=0.48\linewidth]{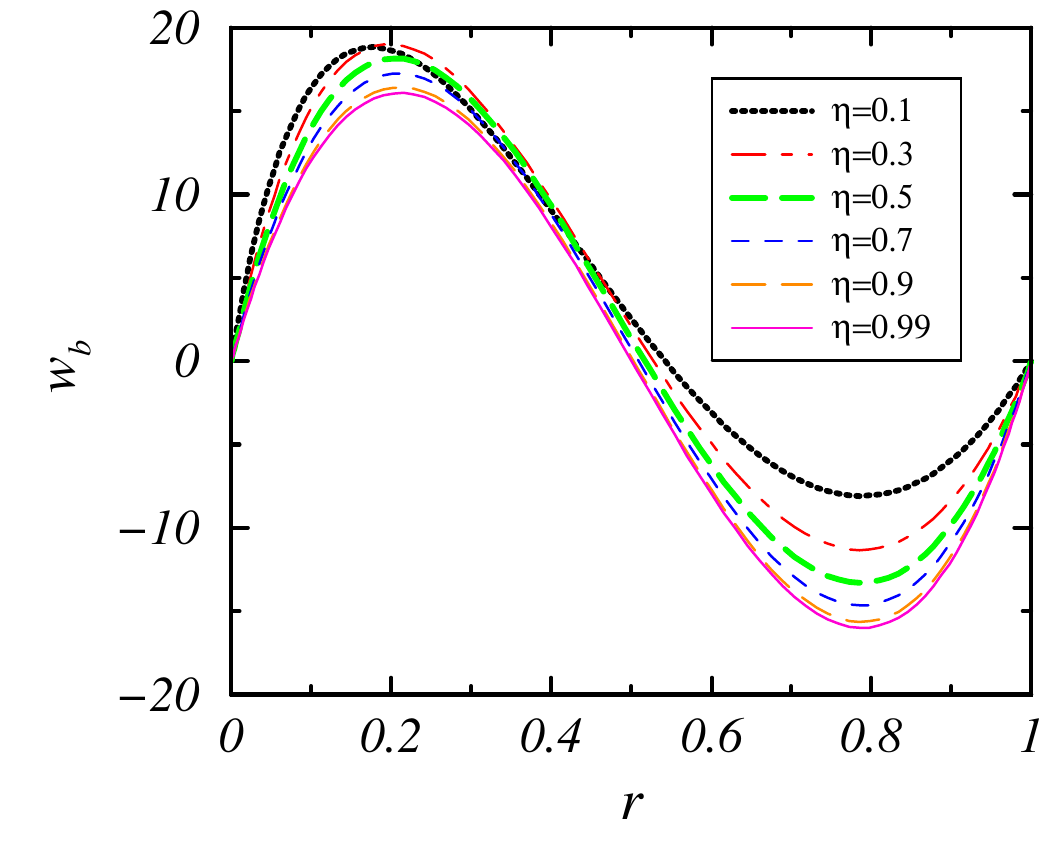}&       
    \includegraphics[width=0.48\linewidth]{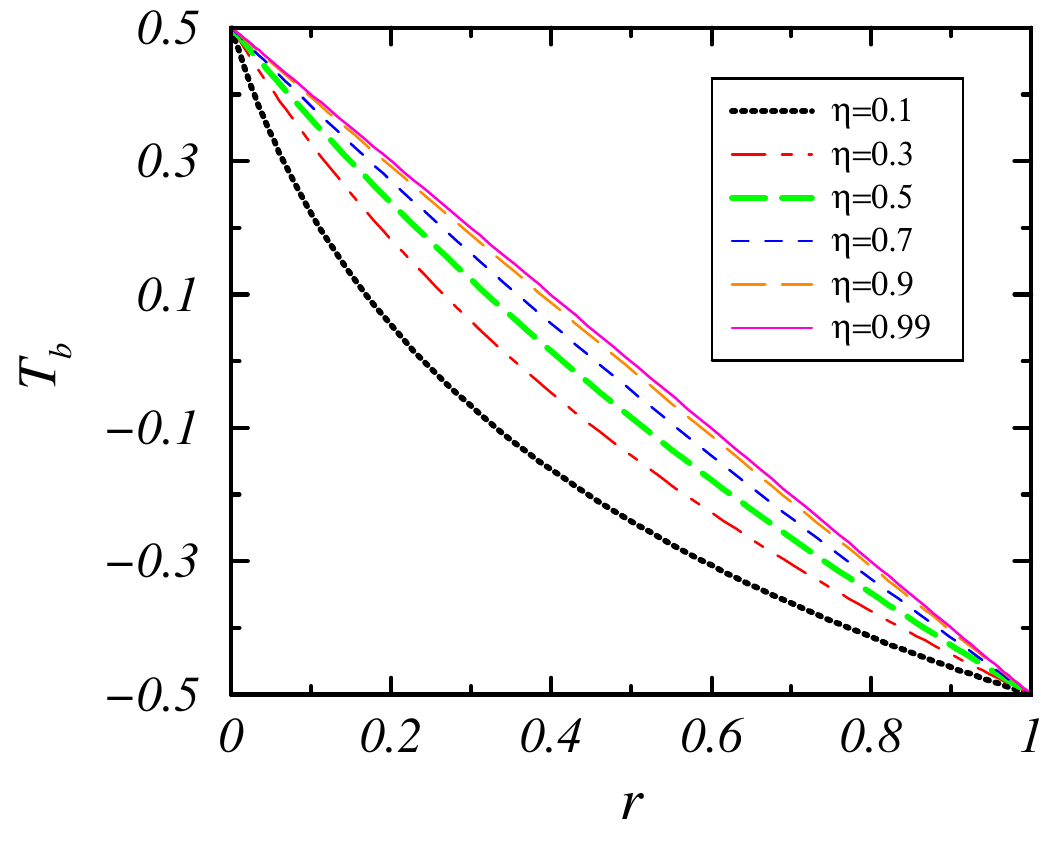}
    \end{tabular}
    \end{center}
\caption{Radial profiles of $(a)$ axial velocity (computed for
  $G=2000$)  and  $(b)$ temperature
  as a function of the radius ratio $\eta$. }
  \label{rad_cond}
\end{figure}

Note that the conductive temperature~\eqref{tem} only depends on 
the radial geometry. Figure~\ref{rad_cond} illustrates 
the variation of the axial velocity and temperature 
with $\eta$.

\subsection{End walls effect: influence of fluid properties and geometry.}

The presence of end walls in the system leads to non-zero radial
velocities close to the end walls, commonly referred to as Ekman
layers, and enables convective heat transfer across the gap even for the basic flow~\citep{Gre68,HiMa75}. 
This switches on the contribution of terms of the heat equation, which were zero under the axial periodicity assumption. 
The temperature equation \eqref{formI_3T} reads
\begin{subequations}\label{Tempeq}
  \begin{align}
    & \partial_t T_b+u_b\partial_rT_b+\frac{v_b}{r}\partial_{\theta}T_b
    +w_b\partial_zT_b=\sigma^{-1}\nabla^2T_b,\label{formII_3T}
  \end{align}
\end{subequations}
In the infinite case the steady basic flow has no radial velocity ($u_b=0$) and the temperature ($T_b$) depends only on $r$, so all terms in the left hand side are zero. Consequently, the basic flow is unaffected by changes in the fluid properties. In contrast, the influence of $\sigma$ may be expected to play a significant role in bounded systems. It modifies the basic flow with respect to the idealised periodic situation and  may become an important source of discrepancies between both systems.  
 
\begin{figure}
\centering
\subfloat[]{
\includegraphics[width=0.45\linewidth]{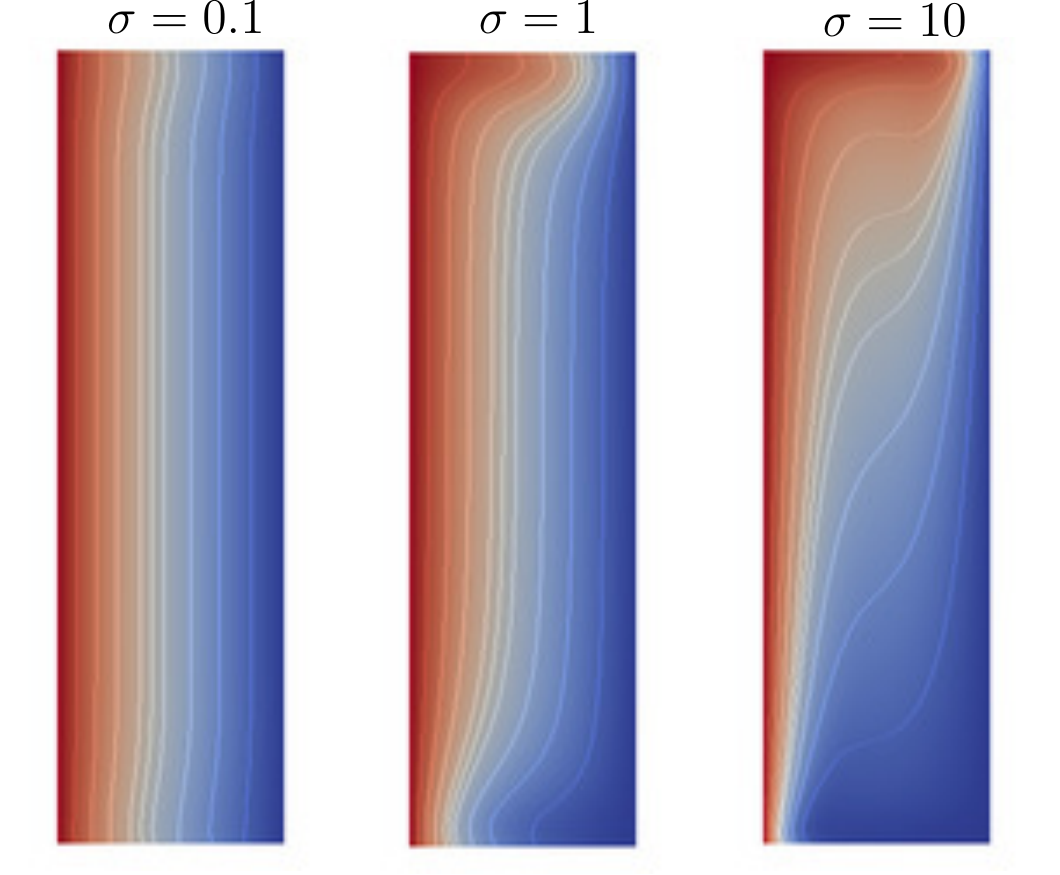}}
\hfill
\subfloat[]{
\includegraphics[width=0.45\linewidth]{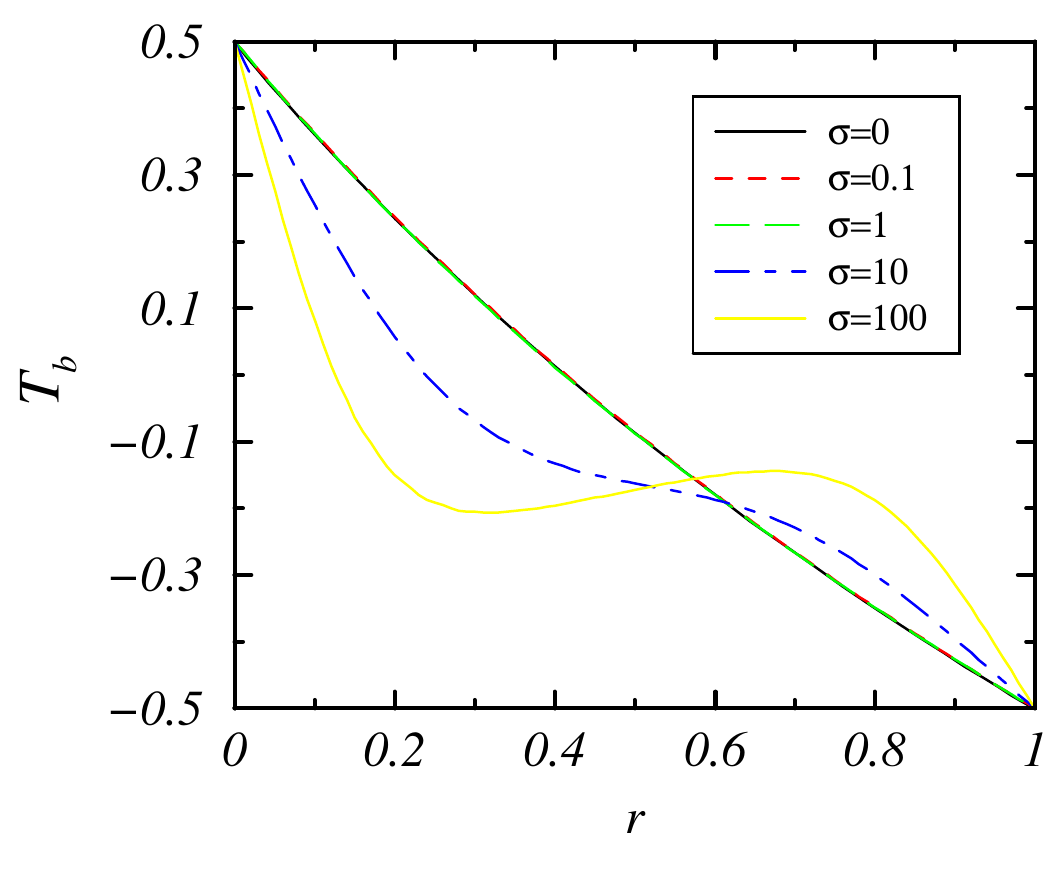}}

\subfloat[]{
\includegraphics[width=0.45\linewidth]{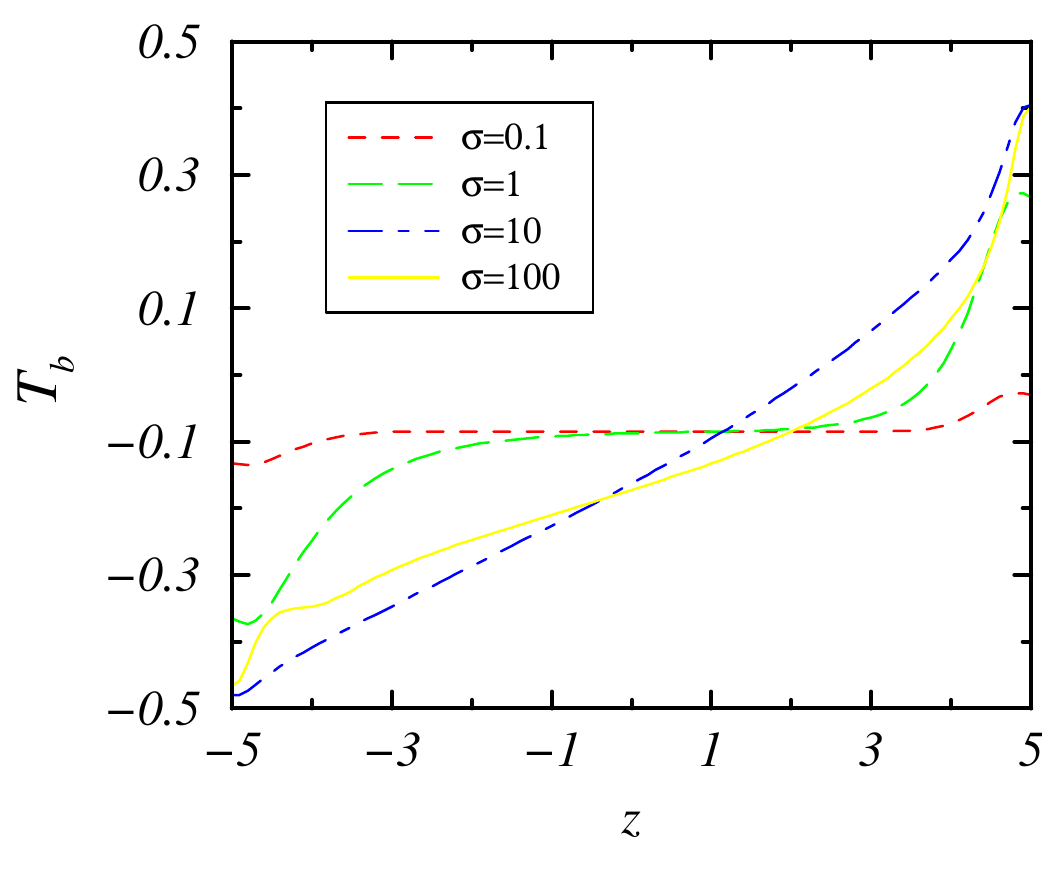}}
\hfill
\subfloat[]{
\includegraphics[width=0.45\linewidth]{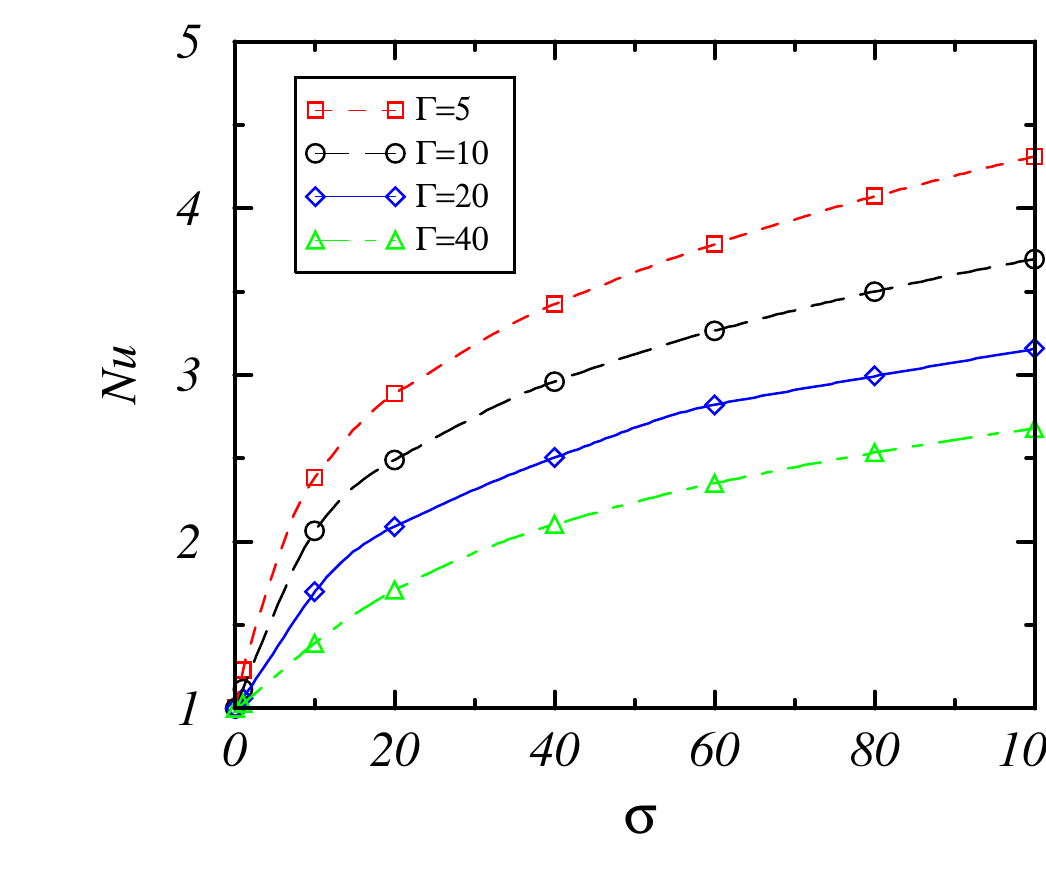}}

    \caption{ (color online) $(a)$ Colormaps of the temperature for
      $\eta=0.5$, $\Rey=30$, $G=2000$, $\Gamma=10$ and
      different values of  $\sigma$. They are plotted in a meridional
      section $(r,z)$ with the inner cylinder on the left hand side.
      Note that the aspect-ratio has been scaled in these plots to enhance the
      visualization of the contours.  (b)--(c) Temperature profiles at
      mid-height and mid-gap, respectively.(d) Variation of $Nu$ with $\sigma$ for several values 
      of  $\Gamma$. The case of $\sigma=0$ has a purely conductive temperature profile as in the axially periodic system. }
    \label{sigma_effect_profiles}
\end{figure}

Figure~\ref{sigma_effect_profiles}($a$) shows colormaps of the
temperature in a meridional section $(r,z)$ for $\eta=0.5$, $\Rey=30$,
$G=2000$, $\Gamma=10$, and $\sigma=0.1, 1, 10$. It clearly illustrates
how changes in $\sigma$ alter the temperature of the basic flow in a
finite system.  Radial and axial temperature profiles are shown
respectively in figures $(b)$ at mid-height and ($c$)  at mid-gap as a
function of $\sigma$. For $\sigma \le 1$ (gases or liquid metals), and
sufficiently far from the end walls, 
the temperature profiles are nearly independent of $z$ and thus
deviations from the conductive regime ($\sigma=0$) are negligible. 
The axial dependence becomes stronger with increasing $\sigma$, 
resulting in radial temperature profiles that differ substantially 
from a conductive profile even at mid-height. 
This temperature regime is usually referred to as
convective~\cite{ThValD70,AlMcF05}, 
and is characterised by sharp thermal boundary layers and 
enhanced heat transfer. 

The heat transfer across the gap is here characterised by the Nusselt of the inner cylinder
\begin{equation}                                                            
  <Nu_i>_t = \frac{\int_0^{2\pi} \int_{-\Gamma/2}^{\Gamma/2}
    \partial_r T|_{r_{i}} \textrm{d}z \textrm{d}\theta}{q_{cond}}      
  \label{eq:Nu}                                                             
\end{equation}                                                              
                                                         
\noindent where $<.>_t$ indicates time average and $q_\text{qcond}$ denotes the dimensionless heat flux of the laminar (conductive) flow in the periodic case. To simplify the notation the subindex $i$ is omitted henceforth. Note that the Nusselt number of the outer cylinder is $Nu_o = Nu_i \eta$. The effect of the transition between the conductive and convective regime on heat transfer is shown in figure~\ref{sigma_effect_profiles}($d$). Large $\sigma$ and short $\Gamma$ promote convective profiles and thus efficient heat transfer.

\subsection{Transition from conductive to convective: the Rayleigh effect}

We quantify the departure of the radial temperature profiles from the conductive regime with the parameter
\begin{equation}
  D = \frac{\int_{r_i}^{r_o} (T_\text{mid}-T_\text{cond})^2}{\int_{r_i}^{r_o}
    T_\text{cond}^2}
  \label{eq:R_par}
\end{equation}
where $T_\text{mid}$ is the radial temperature profile at mid-height and $T_\text{cond}$ denotes the conductive profile~\eqref{tem} of the infinite-cylinder case. 

\begin{figure}
	\begin{center}
		\begin{tabular}{cc}
		  $(a)$   & $(b)$\\                          
		  \includegraphics[width=0.48\linewidth]{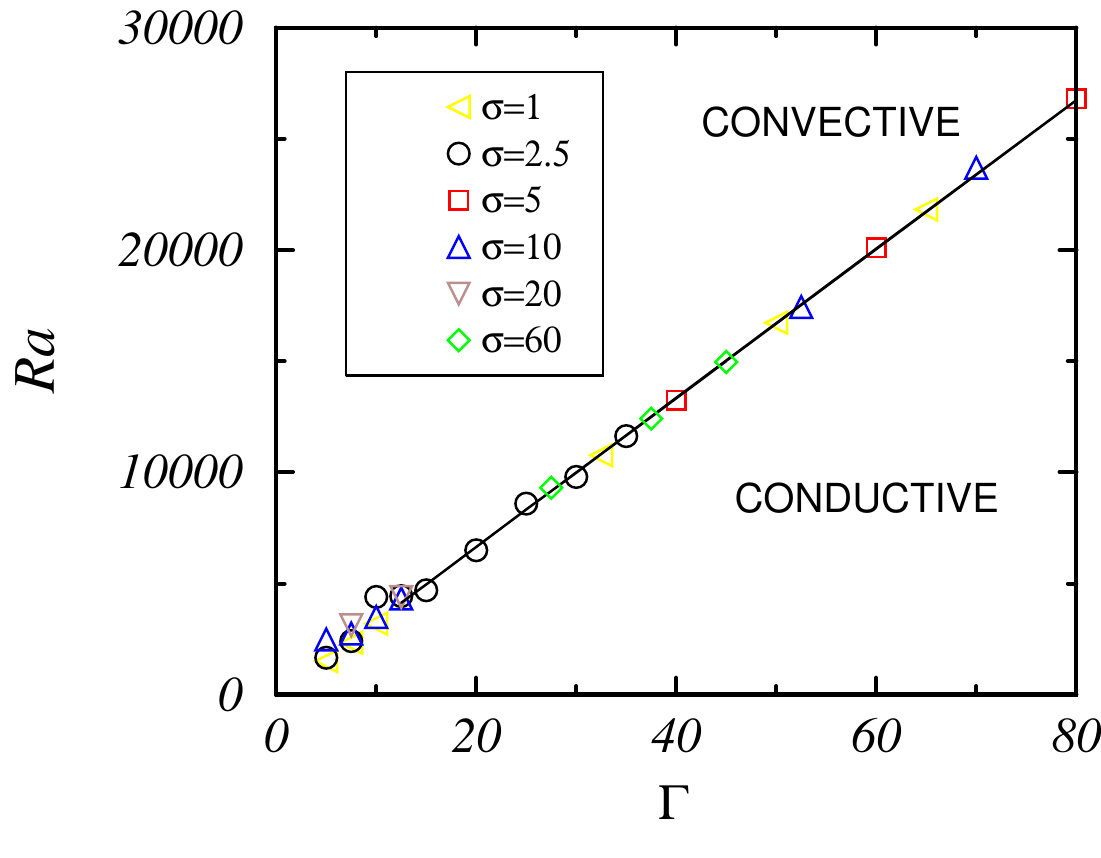}&     
		  \includegraphics[width=0.48\linewidth]{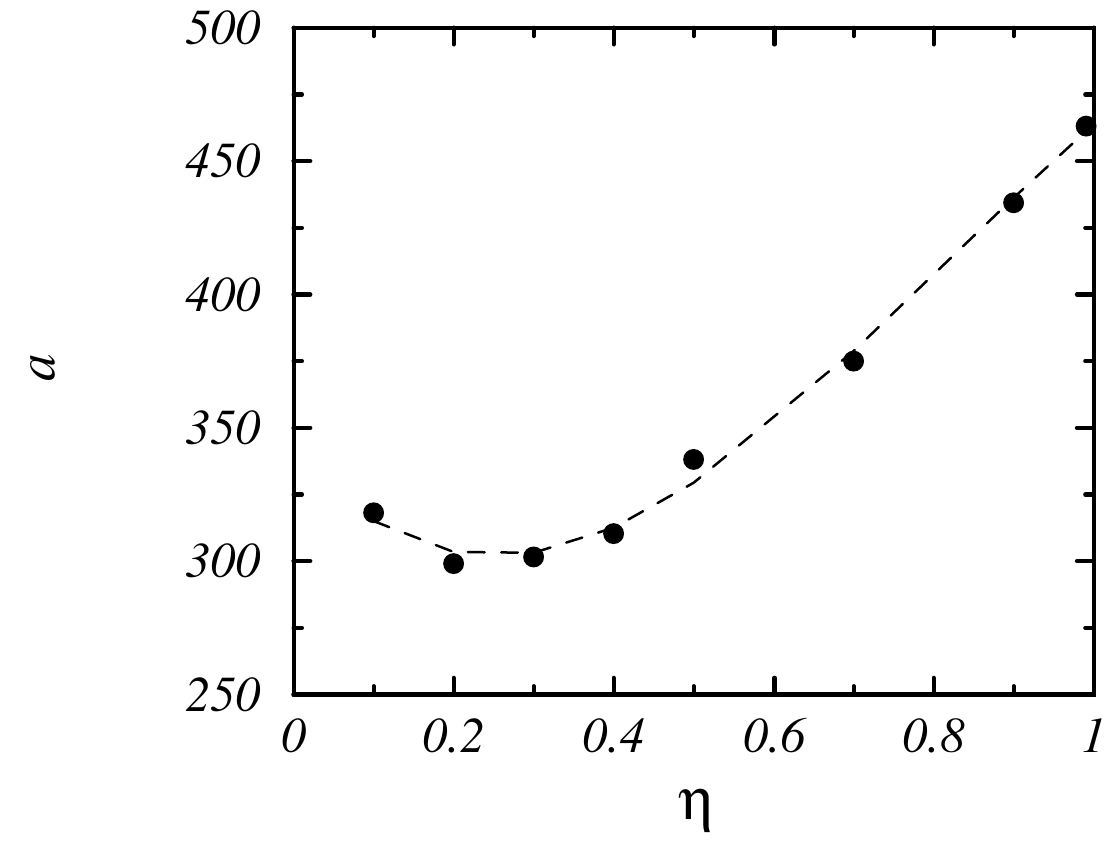}\\	 		
		\end{tabular}
	\end{center}
	\caption{(color online) $(a)$ Transition from conductive to convective temperature regime ($D \gtrsim 10^{-3}$) as a function of the aspect-ratio $\Gamma$ and Rayleigh number $Ra$ for $\eta=0.5$ and $\Rey=30$. $(b)$ Variation of the slope $a(\eta)$ in the linear equation \eqref{eq:crit0.5} marking the transition between the conductive and the convective regimes.} 
        \label{Ra_Gamma}
\end{figure}

Figure~\ref{Ra_Gamma} $(a)$ shows the transition between the conductive and convective regimes as a function of the Rayleigh number ($Ra=\sigma\,G$) and $\Gamma$. The critical Rayleigh numbers $Ra_c$ have been computed by fixing $\sigma$ (using different values) and $\Gamma$, and increasing $G$ until $D \gtrsim 10^{-3}$ was reached. For low values of $\Gamma$ there are significant differences between $Ra_c$ calculated with different $\sigma$. However, for $\Gamma \gtrsim 12.5$ the transition takes place at nearly the same $Ra_c$ regardless of $\sigma$. This interesting behaviour can be explained by analysing equation~\eqref{tem}. The laminar state is steady and axisymmetric, so departures from the conductive regime are due to either $u_b \partial_r T_b$ or $w_b \partial_z T_b$. The first term is only affected by changes in $\sigma$ and plays a significant role in setups with a relatively short aspect-ratio, as the radial velocities arising near the end walls are more vigorous. Consequently, variations of $Ra_c$ with $\sigma$ are observed in short setups. 

In contrast, the second term depends on both $\sigma$ and $G$, but
only through their product: the Rayleigh number $Ra$. This becomes
progressively dominant as $\Gamma$ increases and the transition to the
conductive regime becomes dependent on $Ra$ only. Interestingly,
$Ra_c$ varies linearly with  $\Gamma$ according to
\begin{equation}\label{eq:crit0.5}
  Ra_c(\Gamma)=338.04\Gamma - 216.24
\end{equation}
This linear behaviour was also observed for fixed  $\sigma=1$ by Thomas and De Vhal Davis~\cite{ThValD70}, who studied the transition between both temperature regimes for natural convection.

The linear relationship \eqref{eq:crit0.5} separates the regions of parameter space in which the temperature regime is conductive and conductive. It provides a criterion to determine a priori whether simulations using axially periodic boundary conditions render a good approximation of laboratory flows. Note however that since the temperature profile of the laminar conductive state \eqref{tem} depends on $\eta$, criterion \eqref{eq:crit0.5} is valid only for $\eta=0.5$. In order to investigate the effect of curvature in the transition from conductive to convective heat transfer, we performed simulations as those shown in figure~\ref{Ra_Gamma} $(a)$ for different values of $\eta$. We found that for all investigated curvatures $\eta\in[0.1,0.99]$ a linear relationship of the form
\begin{equation}\label{eq:crit}
  Ra_c(\eta,\Gamma)=a(\eta)\Gamma +b(\eta) 
\end{equation}
is always satisfied, but the slope $a$ depends significantly on $\eta$ (see figure~\ref{Ra_Gamma} $(b)$). The curve $a(\eta)$ is well approximated by the cubic function 
\begin{equation}\label{cubicfit}
  a(\eta) = -307.2\eta^3 + 750.1\eta^2 -318.7\eta + 339.6
\end{equation}
The offset of the linear equation satisfies $b\in [-300,-100]$ for the whole $\eta$-range and so 
can be neglected  in comparison with $a(\eta) \Gamma$ for $\Gamma \gtrsim 15$.

\section{Primary instability}\label{sec:primary}

\begin{figure}
  \begin{center}
    \begin{tabular}{cc}
      $(a)$   & $(b)$  \\
      \includegraphics[width=0.48\linewidth]{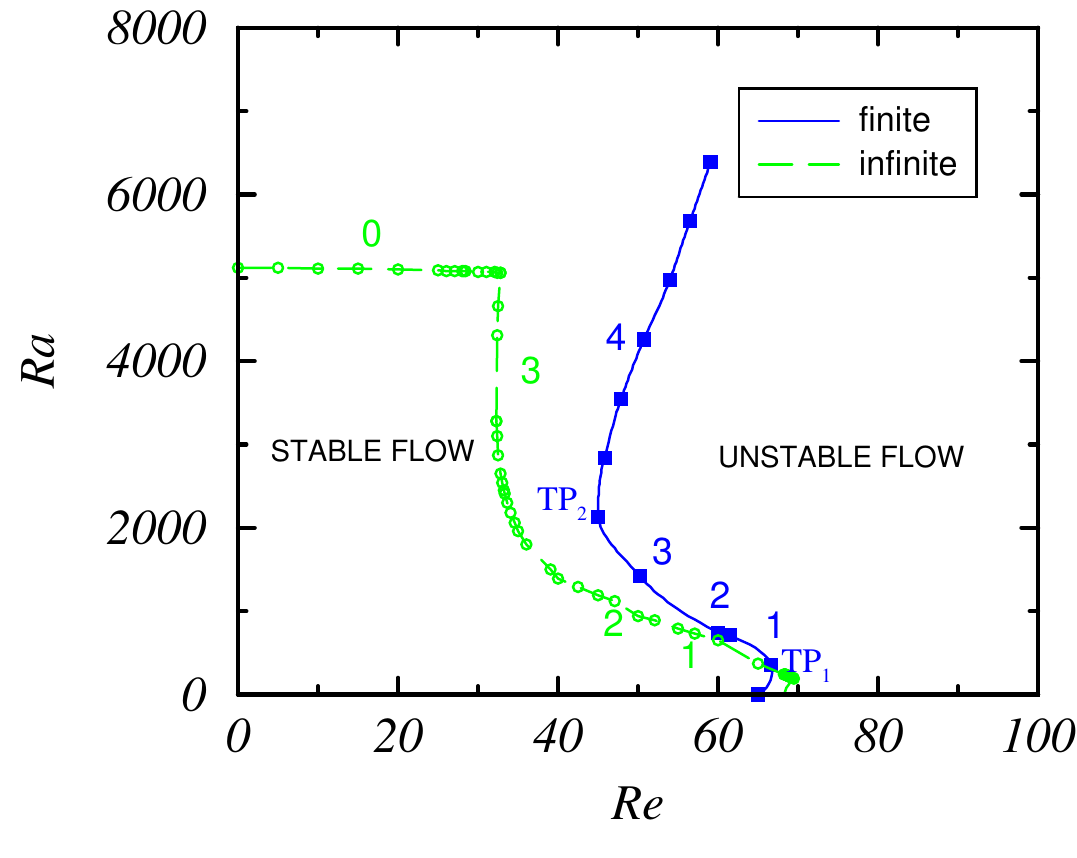}&
      \includegraphics[width=0.48\linewidth]{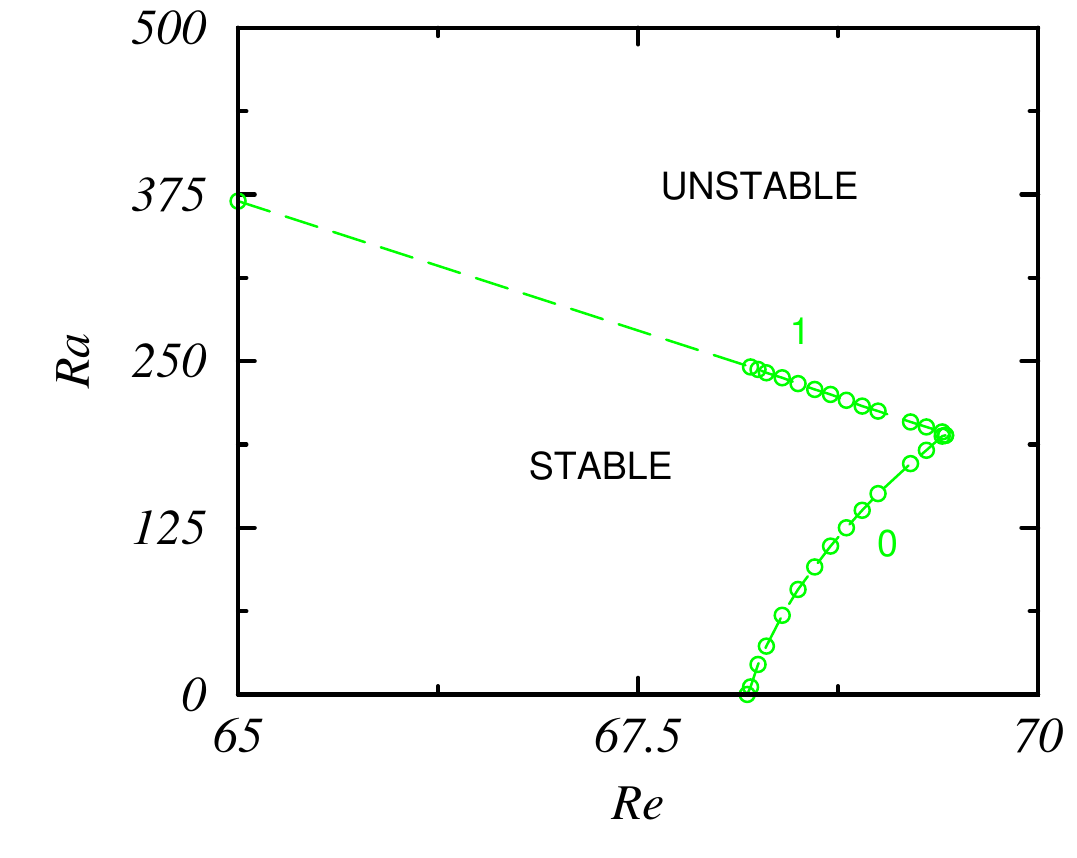}
     
    \end{tabular}
  \end{center}
  \caption{ (color online) $(a)$
    Critical stability boundaries ($\Gr_c$ vs
    $\Rey$). The
    numbers that appear on top of the critical curves are
    the azimuthal modes associated to the critical perturbations.
    (Blue) Solid line and squares are used in the case of 
    finite-length cylinders, whereas (green) dashed line and circles 
    stand for the marginal stability in the case of infinite 
    long cylinders. The working
    fluid is air ($\sigma=0.71$) and the geometric parameters are
    $\eta=0.5$ and  $\Gamma=10$ (in the finite case).
    $(b)$ Detail of the marginal curve for low $\Gr$ in the	
		infinite cylinders case.}
    \label{stability}
\end{figure} 

\subsection{Axially periodic disturbances}\label{stab:pbc}

The dashed line in figure~\ref{stability} $(a)$ shows the stability 
curve of the conductive laminar flow to axially periodic disturbances
for $\eta=0.5$ and $\sigma=0.71$. Three regions with qualitatively different behaviour can be
distinguished.  For slow or no rotation the laminar flow is destabilised 
as $Ra$ increases. Buoyancy is the indirect force driving the instability, which occurs because the axial velocity profile~\eqref{uz} has an inflection point. At low $Ra$ viscosity can stabilise the laminar flow, but as $Ra$ increases the magnitude of the axial velocity grows proportionally to $Ra$ until the flow becomes unstable at $Ra_c=5120$. The instability pattern consists of 
axisymmetric convection rolls that travel axially in the upward direction~\cite{VaTh69}. This axial drift of the rolls is caused by curvature, which makes the axial velocity profile of the conductive basic flow radially asymmetric (see Fig.~\ref{rad_cond}). In the limit of vanishing curvature $\eta\rightarrow 1$ (vertical plates) the roll-pattern is steady because the system is left-right symmetric about the mid-gap~\cite{elder1965laminar}. 

As the inner-cylinder rotation speed increases the centrifugal force starts to compete with 
buoyancy and the laminar flow becomes first unstable to
nonaxisymmetric spiral modes. Although centrifugal instabilities are
typically axisymmetric (i.e~Taylor vortices), it is well known that in
the presence of a sufficiently large axial velocity nonaxisymmetric
spiral modes are dominant. This was first shown by
Snyder~\cite{snyder1962experiments} in Taylor--Couette flow with a
superimposed axial pressure gradient. In our system, as $Re$ increases
the critical Rayleigh number for the onset of these spiral modes and
their wavenumber ($n$) progressively decrease. At
  $(\Rey,Ra)=(69.42,194.53)$ the critical curve for the spiral modes
  intersects the curve corresponding to the pure centrifugal
  instability (i.e~Taylor vortices with $n=0$). 
Figure~\ref{stability} $(b)$ shows a close up to the vicinity of this point. 

\subsection{Axially bounded cylinders} 

Following Kuo and Ball~\cite{KuBa97} we chose $\Gamma=10$.  
The computed stability curve is shown as a solid line 
in figure~\ref{stability} $(a)$ and also features an intersection
point around which spiral modes and axisymmetric Taylor
vortices compete.  In comparison to the periodic case, this intersection point 
is displaced to $ (\Rey,Ra) = (66,257.02)$ because of the
finite aspect-ratio. Here, as in the periodic case, and 
long aspect-ratio experiments of  Snyder and Karlson~\cite{SK64}, 
who used $\Gamma= 337$, there is a region in which spiral modes
appear at lower values of $\Rey$ than Taylor vortices. 
However, this region is limited by a turning point and when $Ra$ is further increased
instabilities are suppressed. The Ekman layers cause a slow
transition to the convective basic state and enhance heat transfer
without the need for instabilities. Overall the stability curves for 
the finite and periodic system are in qualitative agreement up to 
$Ra\approx 3000$. This is in line with criterion \eqref{eq:crit0.5}, which predicts the 
transition between conductive and convective states  at $Ra = 3164$. 

\section{Heat transfer for increasing rotation speed}\label{sec:bif_fixT}

The stability analysis of the previous section cannot predict the
dynamics and heat transfer properties of the flow. 
For these purposes we performed direct numerical simulations of the
Navier--Stokes equations. We first investigated $Ra=1420$, 
so that the temperature for $\Gamma=10$ is conductive. 
In the periodic case we choose a cylinder length of $L_z=7.48$, 
which allows some freedom in the selection of the wavenumber of the vortex-pairs. 

\begin{figure}
   \begin{center}
     \begin{tabular}{ccccc}
       $(a)$   & $(b)$  & $(c)$  &
       $(d)$ & $(e)$ \\

         $\Rey=50$   &  $\Rey=70$  &  $\Rey=135$ &
        $\Rey=260$ &  $\Rey=300$\\
       \includegraphics[scale=0.25]{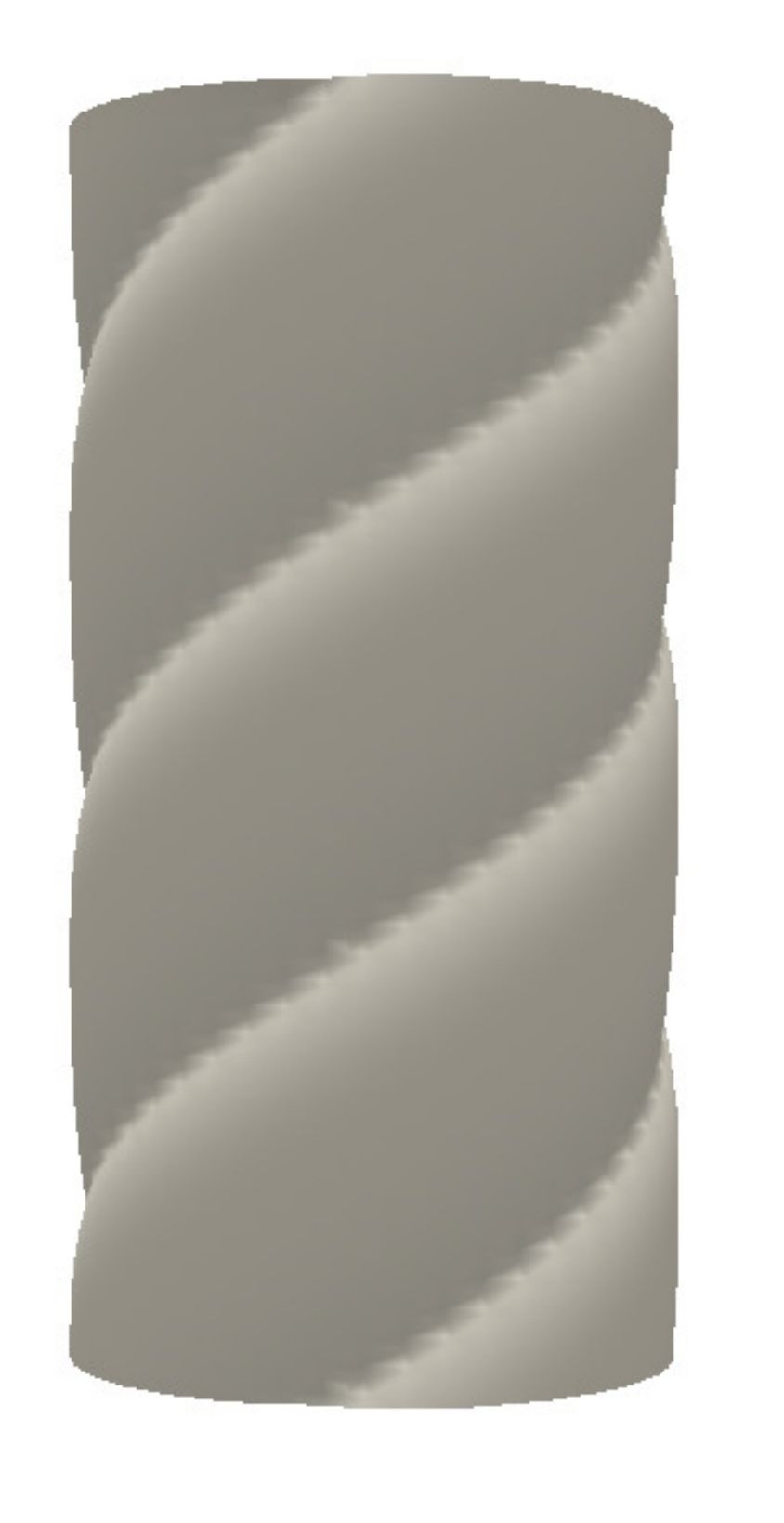}&
       \includegraphics[scale=0.25]{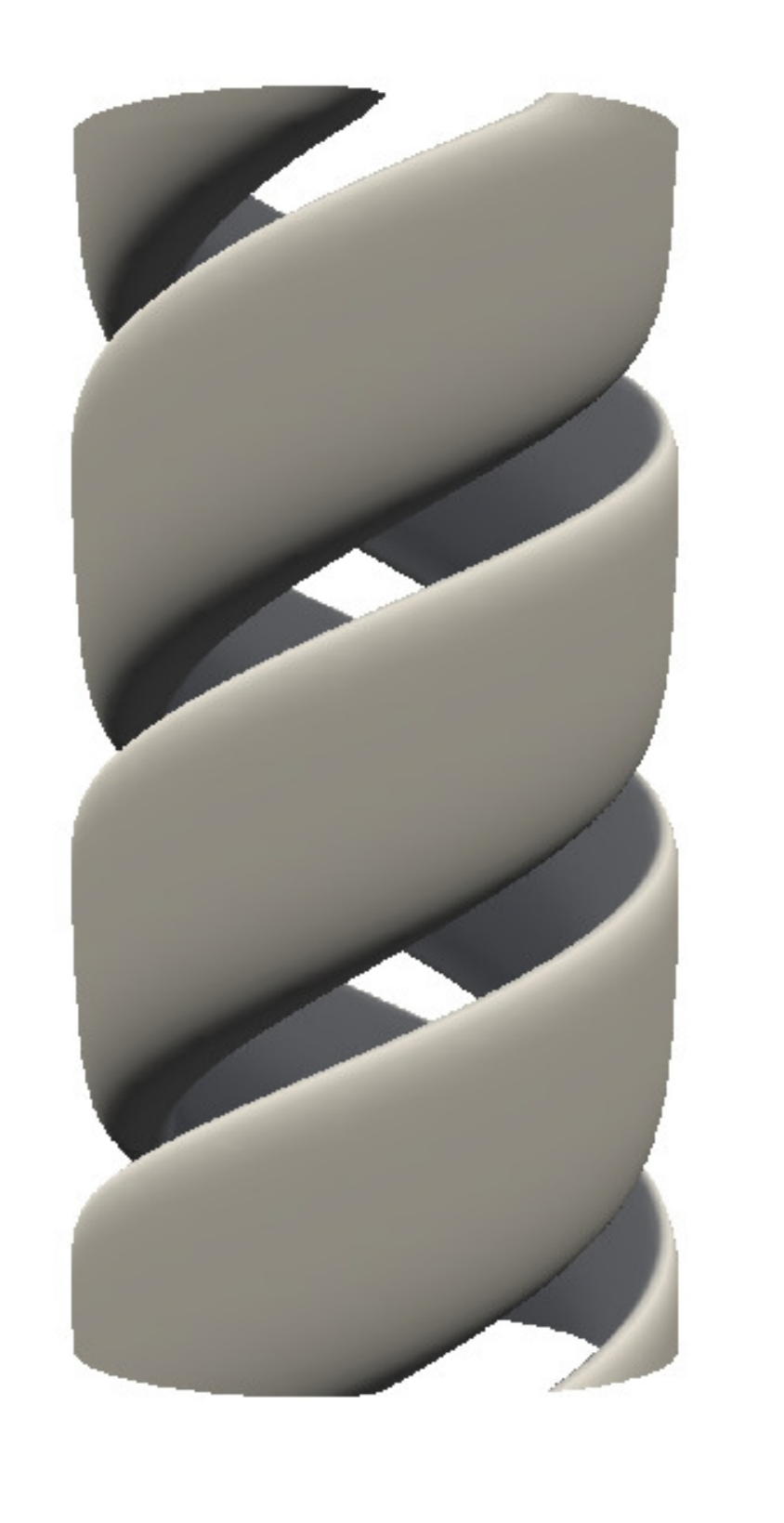}&
       \includegraphics[scale=0.25]{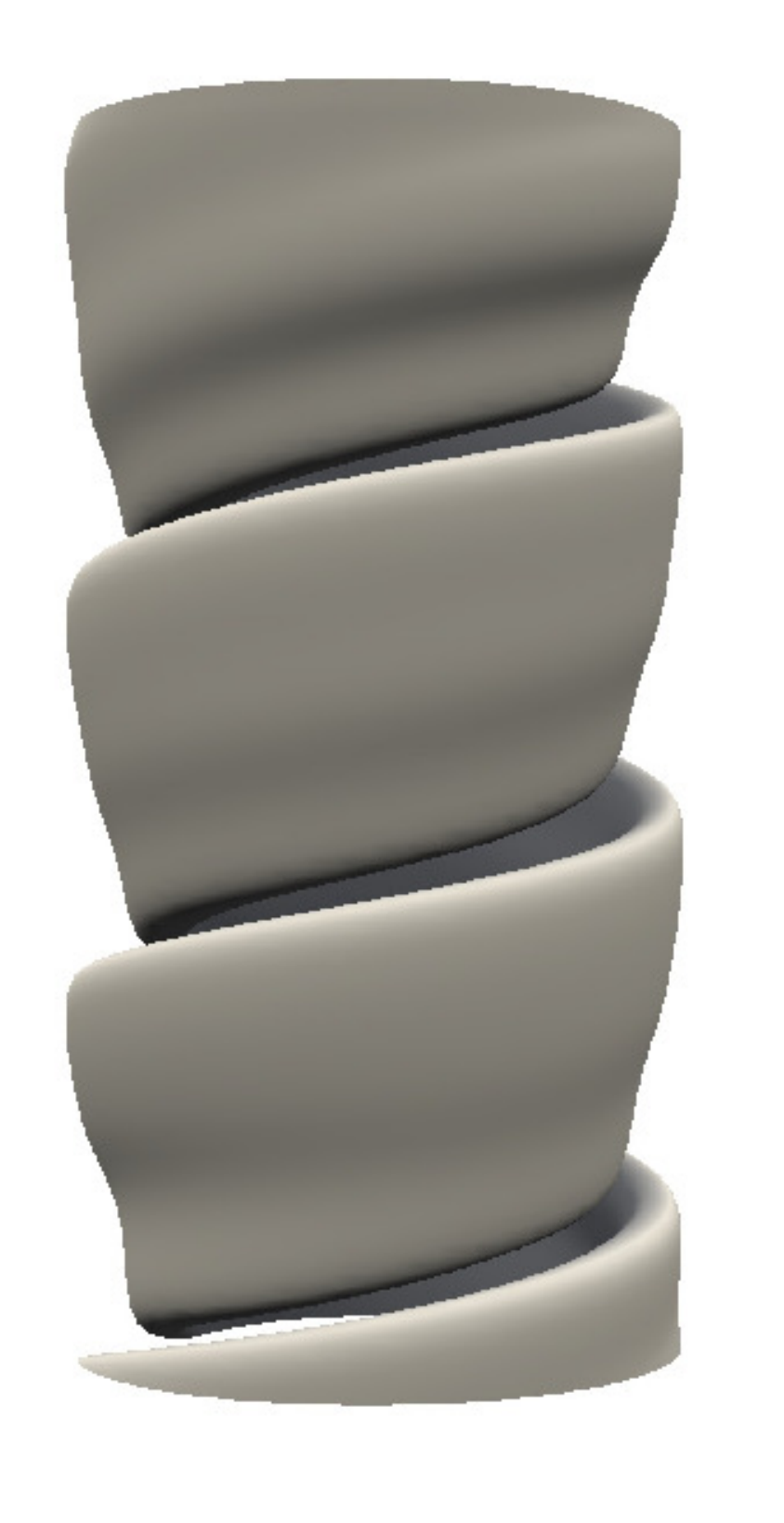}&
       \includegraphics[scale=0.25]{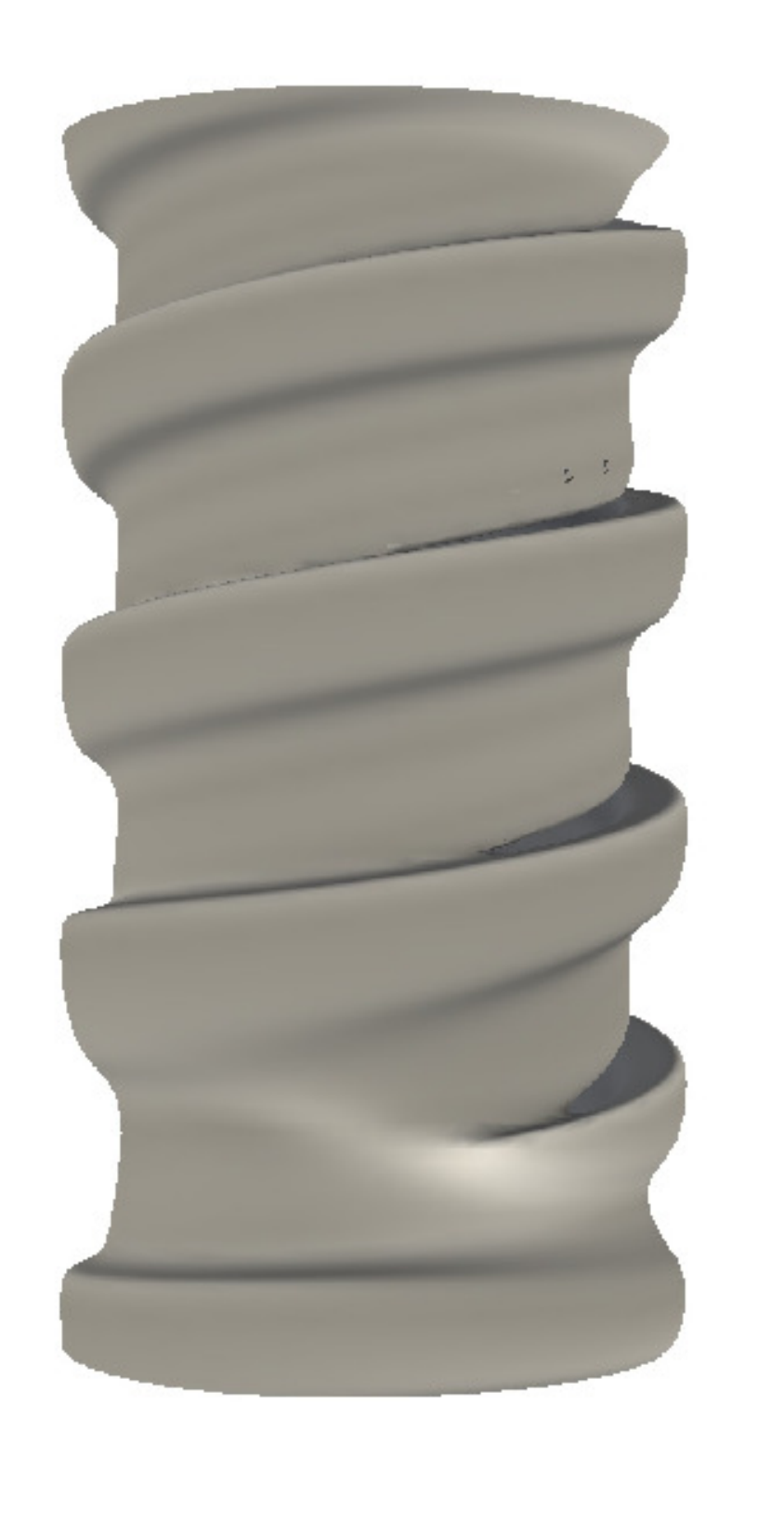}&
       \includegraphics[scale=0.25]{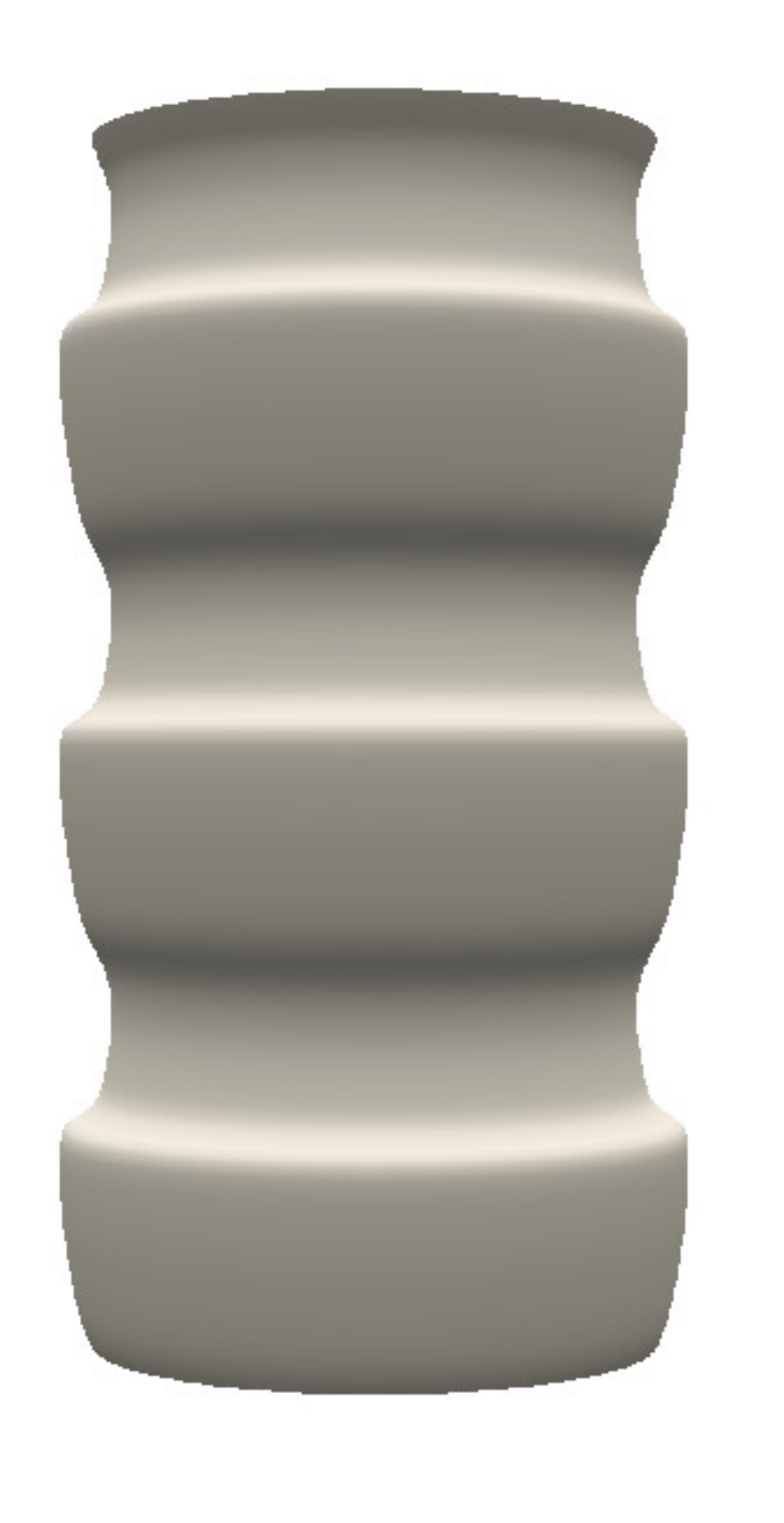}\\
       $(f)$   & $(g)$  & $(h)$  &
       $(i)$ & $(j)$ \\
       $\Rey=60$  &  $\Rey=70$  &  $\Rey=90$ & 
       $\Rey=120$ & $\Rey=140$\\
       
       \includegraphics[scale=0.25]{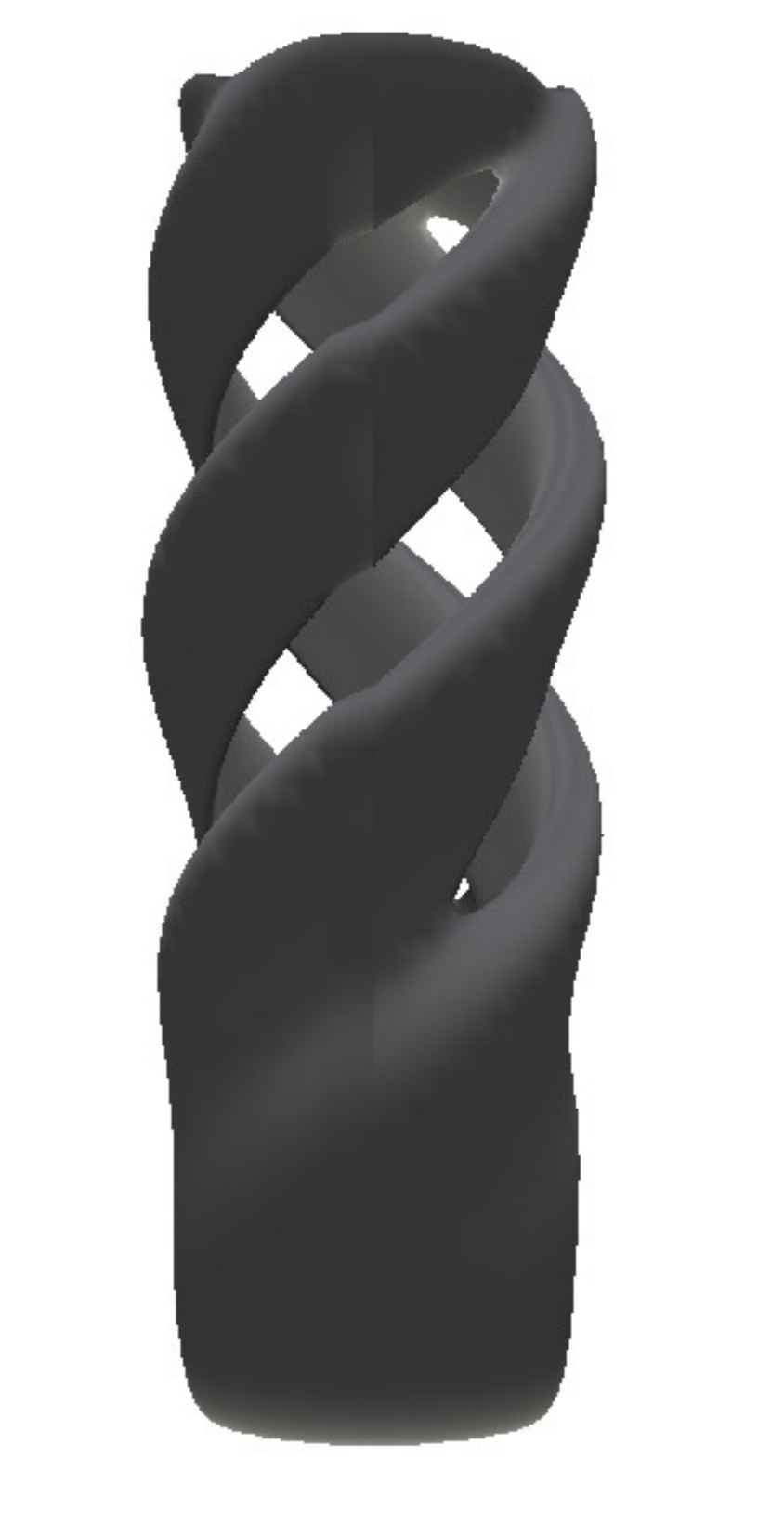}&
       \includegraphics[scale=0.25]{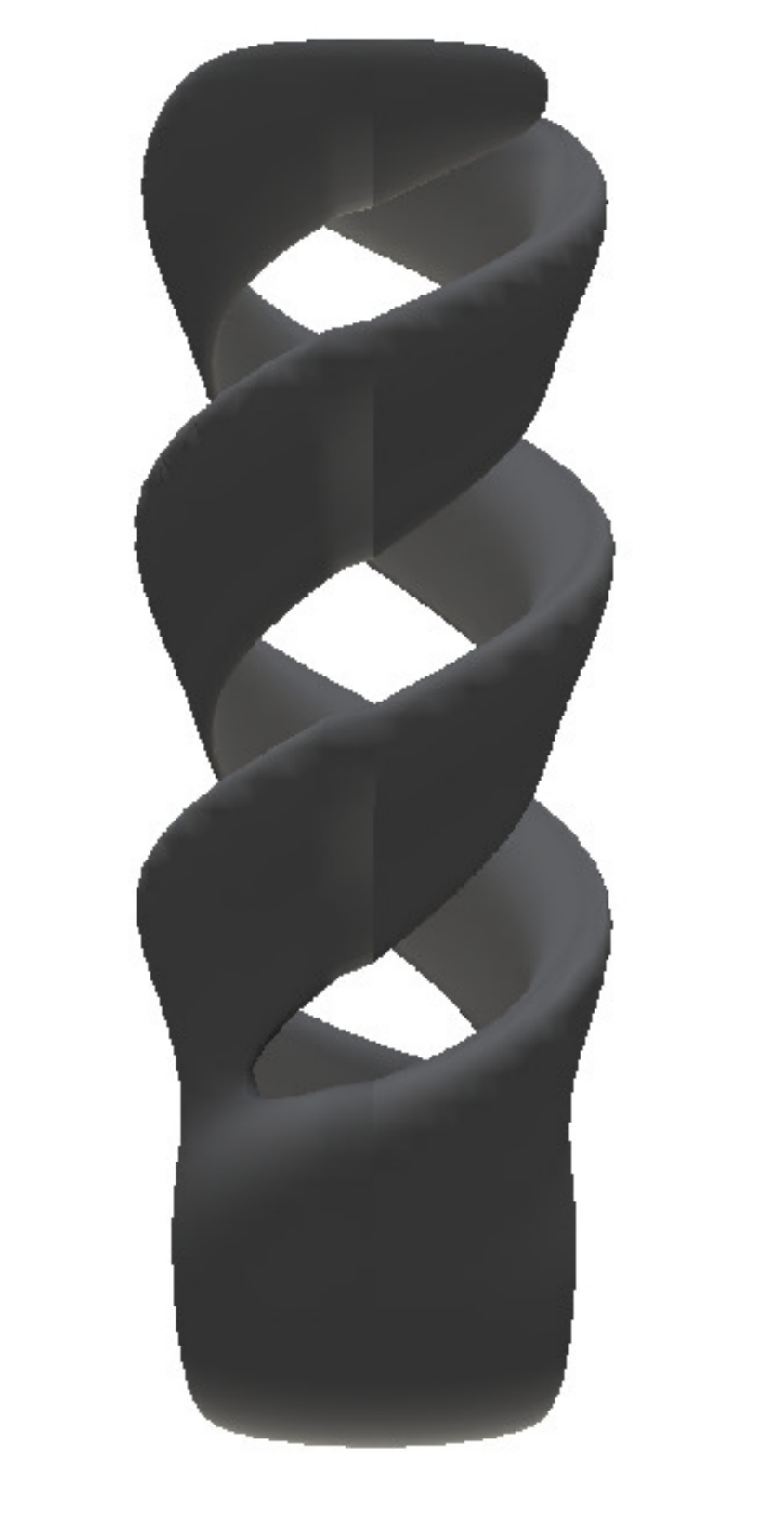}&
       \includegraphics[scale=0.25]{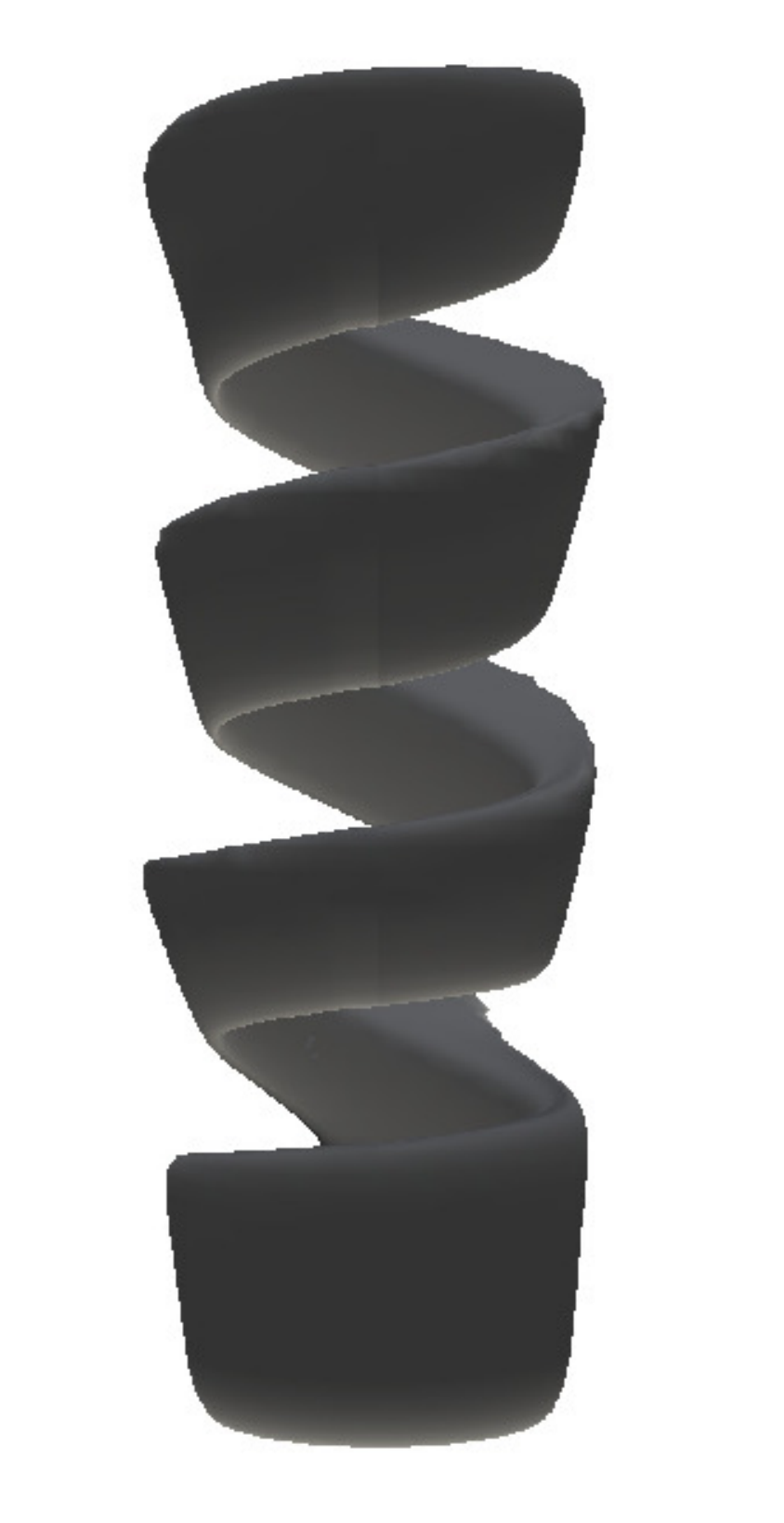}&
       \includegraphics[scale=0.25]{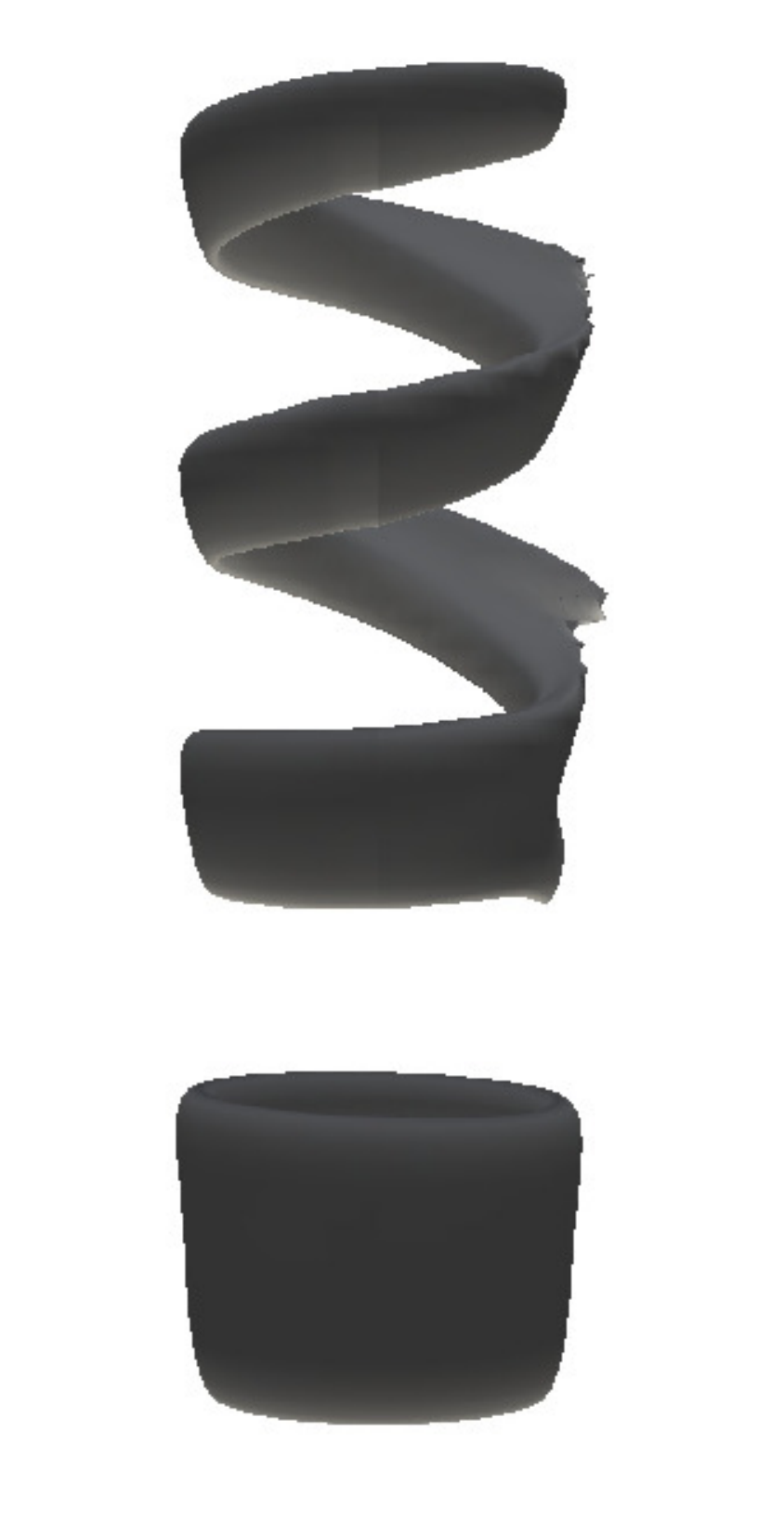}&
       \includegraphics[scale=0.25]{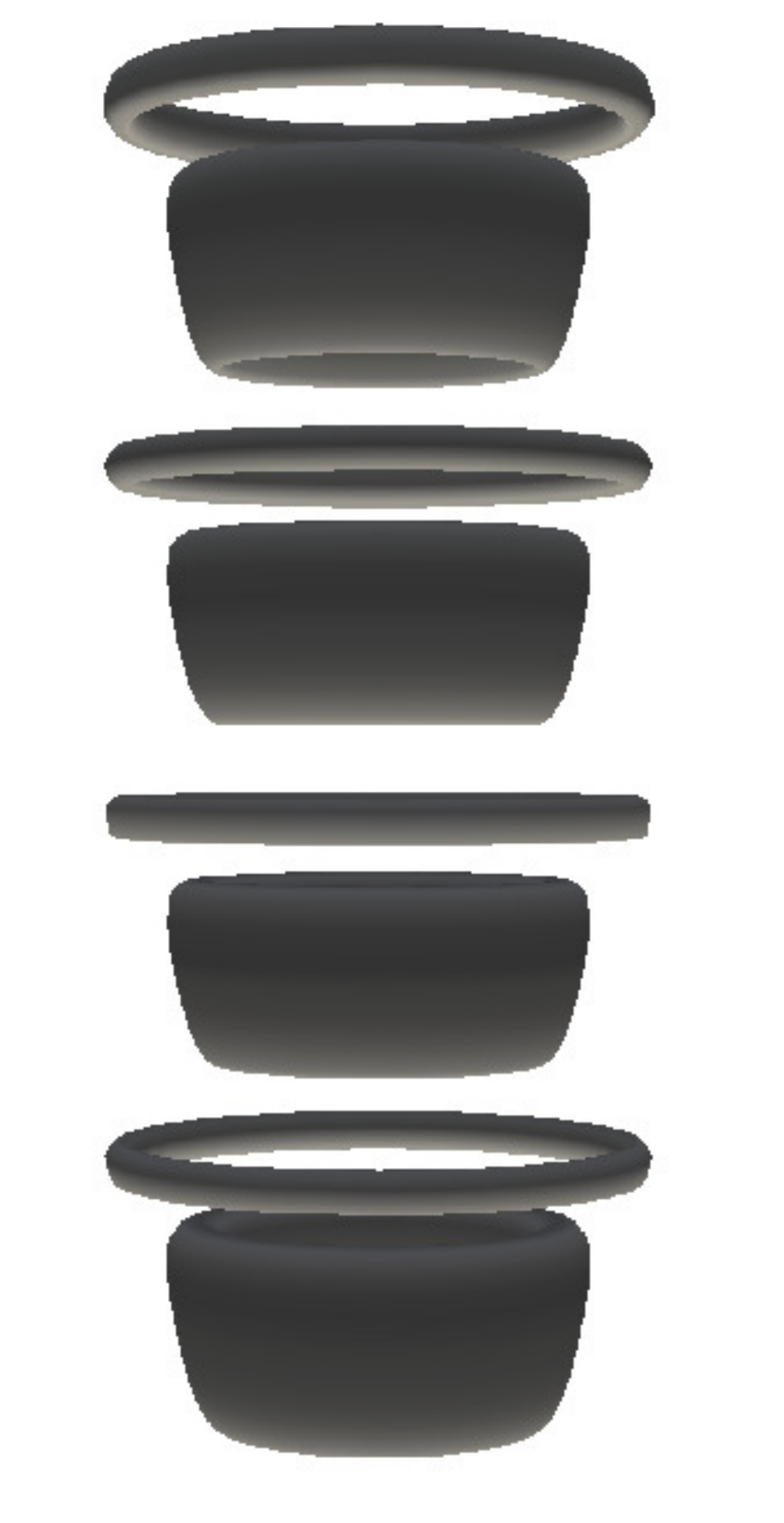}
     \end{tabular}
   \end{center}
   \caption{ Evolution of an isosurface of the axial velocity
     in the periodic (upper row) and finite (lower row) systems 
     as $\Rey$ is varied and $Ra=1420$, $\sigma=0.71$, $\eta=0.5$,$\Gamma=10$
     (finite case) and $L_z=7.48$ (infinite
     case).}\label{reg3_2}
 \end{figure}

Figure~\ref{reg3_2} shows the sequence of emerging patterns in the
periodic (upper row) and finite (lower row) cases as the Reynolds number is
increased. For the periodic system the primary instability occurs at $\Rey \approx 39$, leading to spiral flow 
with azimuthal wavenumber $n=3$ (figure~\ref{reg3_2} $(a)$). By further increasing $\Rey$ the axial flow  generated by buoyancy loses importance in comparison to the centrifugal force. This results in secondary transitions towards spiral flows with decreasing azimuthal mode $n$. The transition from $n=3$ to $n=2$ takes place at $\Rey \approx 52$,  whereas the transition to $n=1$ occurs at $\Rey
\approx 135$. The flow patterns with $n=2$ and $n=1$ are shown in 
figure~\ref{reg3_2} $(b)$ and $(c)$, respectively. At $\Rey \approx 260$ 
the spiral flow is modulated by a low frequency and the flow becomes quasi-periodic. 
The resulting state preserves the spiral structure  with $n=1$ and is
characterised by the appearance of spatio-temporal defects
(figure~\ref{reg3_2} $(d)$).  Finally,  for $\Rey \geq
    280$ the centrifugal force completely dominates and 
spiral flow is superseded by Taylor vortices
(figure~\ref{reg3_2}~$(e)$). 
These results are consistent with the physical mechanisms discussed 
in the linear stability analysis of \S\ref{stab:pbc}.

The same sequence of flow states was found in the finite case, 
but with slightly different Reynolds number for the transitions. 
The basic flow loses stability to spiral flow with $n=3$ for $\Rey
\approx 50$ (figure \ref{reg3_2} $(f)$), 
whereas the transitions to $n=2$ and $n=1$ occur at $\Rey \approx 63$
and $\Rey \approx 90$ respectively (figure \ref{reg3_2} $(g)$--$(h)$)). 
The transition between spiral flow and Taylor vortices is also
characterised by the presence of defects, 
which are much more pronounced than in the infinite cylinders case
(figure~\ref{reg3_2} $(i)$). 
The onset of Taylor vortices (figure~\ref{reg3_2} $(j)$) 
takes place at $\Rey \approx 130$.

\begin{figure}
   \begin{center}
     \includegraphics[width=0.58\linewidth]{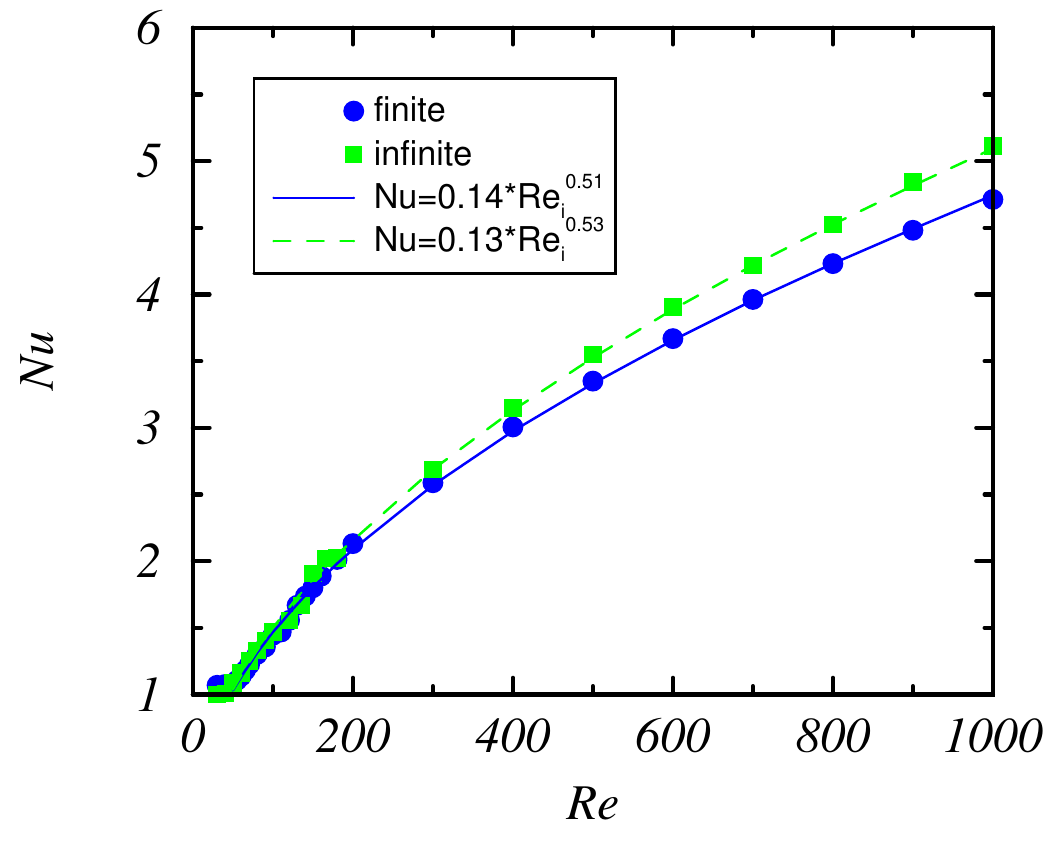}
     \end{center}
   \caption{Variation of $Nu$ with $\Rey$ for $Ra=1420$,
     $\sigma=0.71$ and $\eta=0.5$ in the finite ($\Gamma=10$) and
     infinite ($L_z=7.48$) cases.}
   \label{Nuss_Re}
 \end{figure}

In figure~\ref{Nuss_Re} we show the heat transfer rate, normalised by
the conductive rate as in equation \eqref{eq:Nu}, 
as a function of increasing $\Rey$. 
As the basic flow is conductive for periodic and finite boundary
conditions, $Nu$ is in both cases independent of $\Rey$ before the
onset of instabilities ($\Rey \lesssim 50$). 
In the infinite case heat transfer is purely conductive ($Nu=1$),
whereas in the finite case,  $Nu$ is slightly higher ($Nu=1.07$) due to 
the heat transfer at the Ekman boundary layers. 
The onset of instability enhances the convective heat transfer
significantly and the Nusselt number follows a power-law scaling
$Nu=A\Rey^B$, with $(A,B)=(0.14,0.51)$ for the finite case and
$(A,B)=(0.13,0.53)$ for the periodic case. 

\section{Heat transfer for increasing temperature difference}\label{sec:bif_fixrot}

\begin{figure}
   \begin{center}
     \begin{tabular}{cccc}
       $(a)$ & $(b)$  & $(c)$  & $(d)$\\
       $Ra=2130$ & $Ra=2840$ & $Ra=3550$ &  $Ra=4970$\\
       \includegraphics[width=0.23\linewidth,height=4cm]{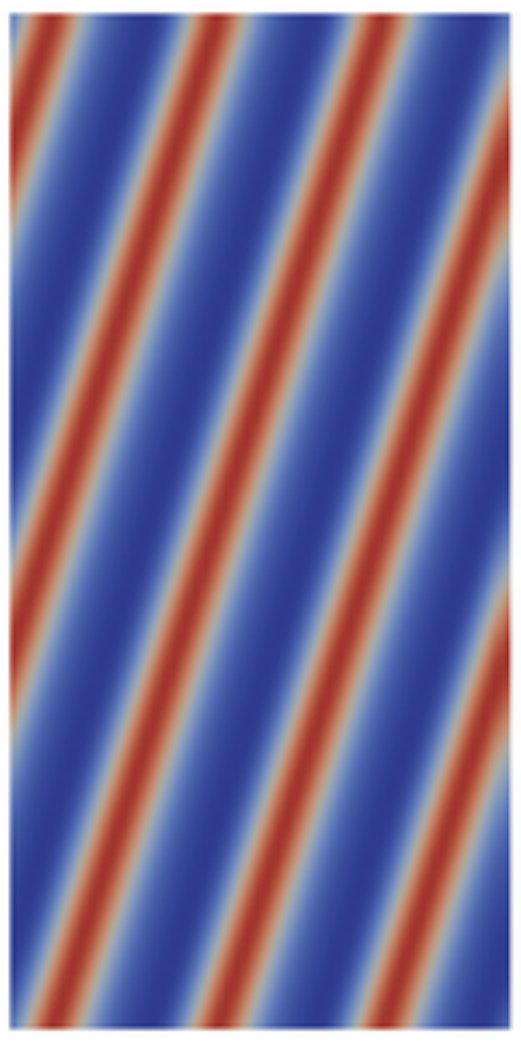}&                                                                    
       \includegraphics[width=0.23\linewidth,height=4cm]{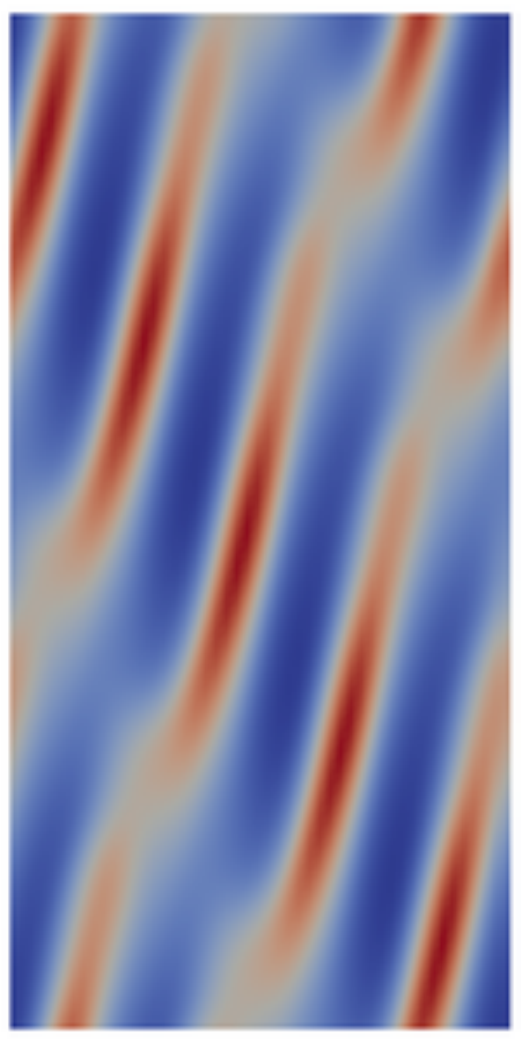}&                                                                    
       \includegraphics[width=0.23\linewidth,height=4cm]{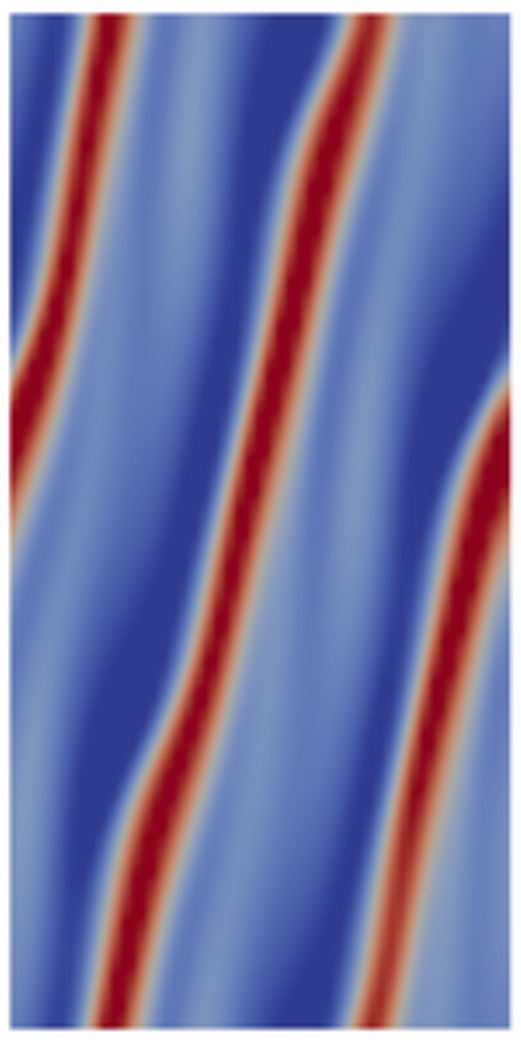}&                                                                    
       \includegraphics[width=0.23\linewidth,height=4cm]{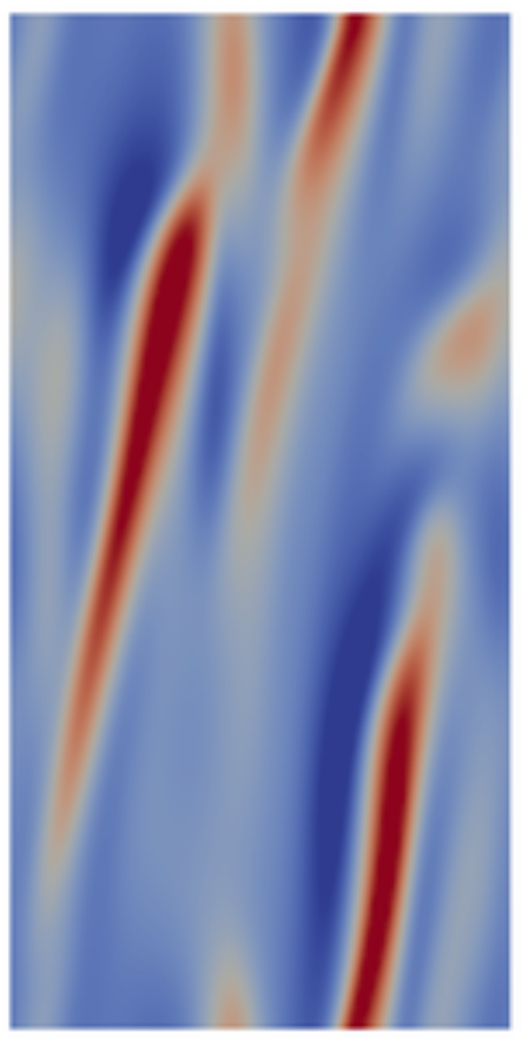}\\ 
       $(e)$ & $(f)$  & $(g)$  & $(h)$\\
        $Ra=1420$ &  $Ra=1775$ &  $Ra=3550$ &  $Ra=5680$\\
       \includegraphics[width=0.23\linewidth,height=4cm]{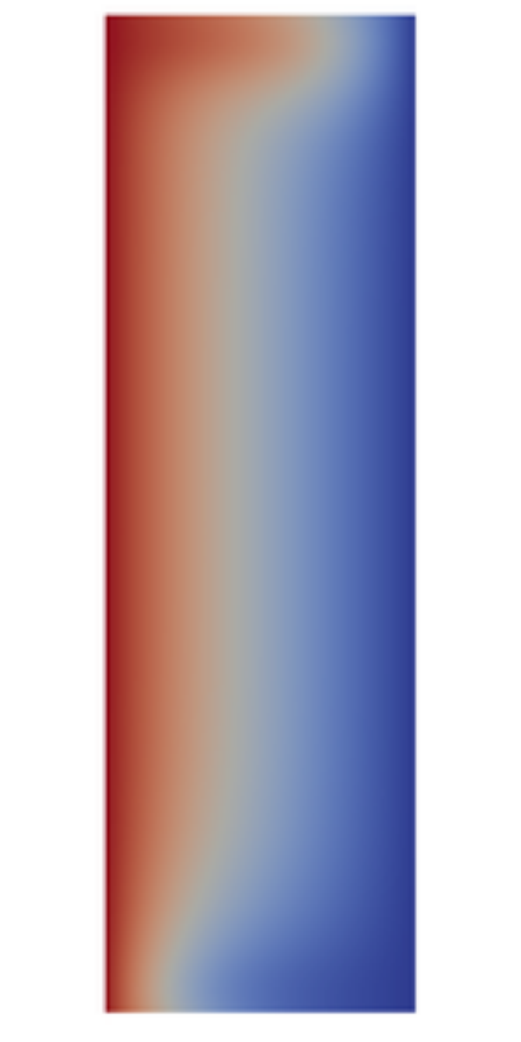}&
       \includegraphics[width=0.23\linewidth,height=4cm]{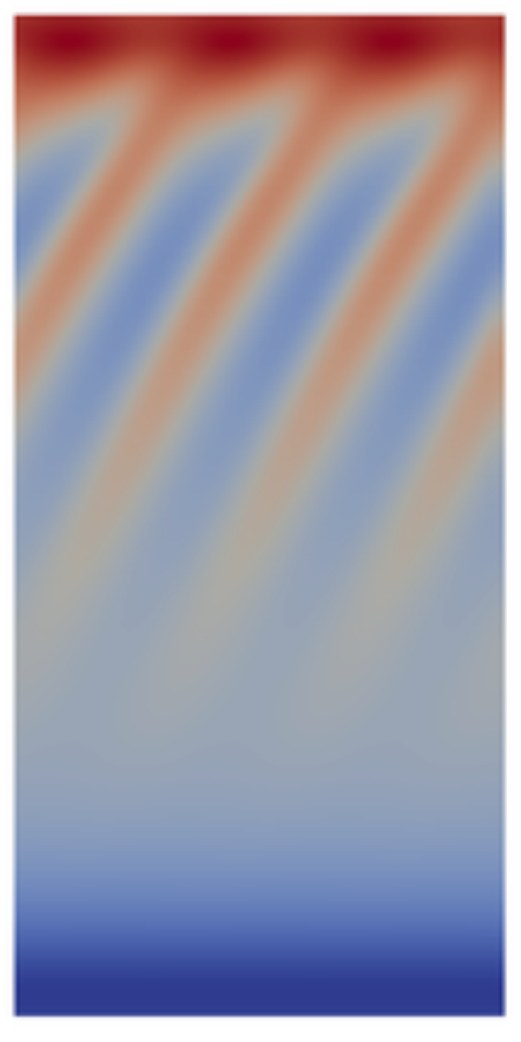}&
       \includegraphics[width=0.23\linewidth,height=4cm]{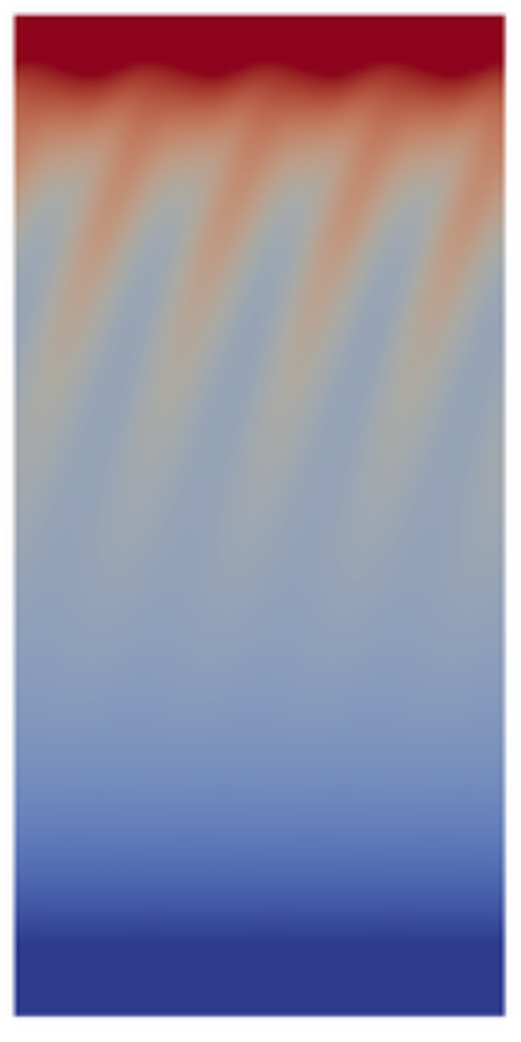}&
       \includegraphics[width=0.23\linewidth,height=4cm]{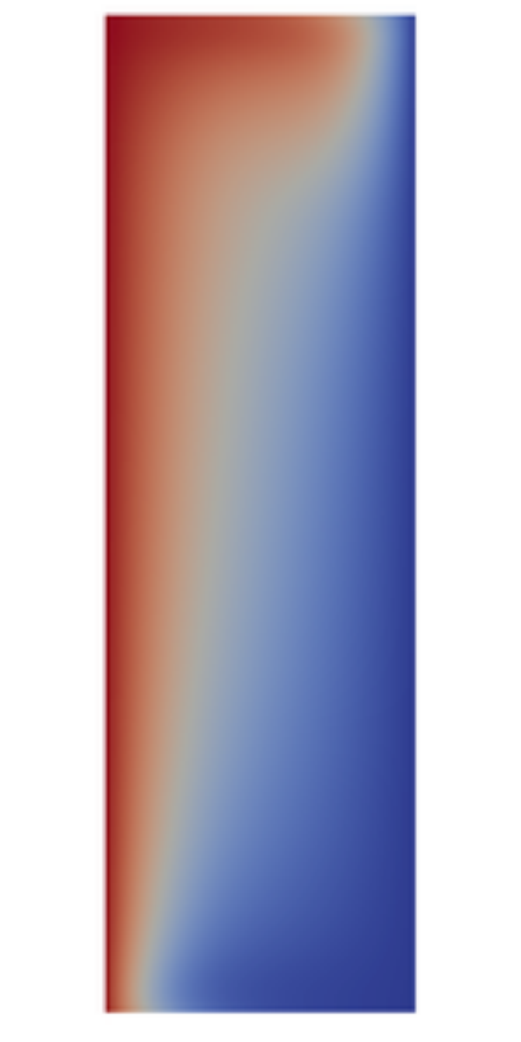}
 \end{tabular}
   \end{center}
   \caption{ (color online) Evolution of the temperature 
     in the periodic (upper row) and finite (lower row) 
     systems as $Ra$ changes and $\Rey=50$, $\sigma=0.71$, $\eta=0.5$,
     $\Gamma=10$ (finite case) and $L_z=7.48$ (infinite
     case). Colormaps
     are plotted in longitudinal sections $(\theta,z)$  at
     $r=\frac{r_o+r_i}{2}$, with the exception of the axisymmetric 
     states $(e)$ and $(h)$, which  are plotted in a meridional
     section $(r,z)$. Note that these plots have been resized and do not 
     correspond to the aspect-ratio of the investigated system.}
   \label{Ra_increases}
 \end{figure}

We here investigate the dynamics and heat transfer for $\Rey=50$ and
increasing $Ra$. The upper row of figure~\ref{Ra_increases} illustrates
the bifurcation scenario for axially periodic boundary conditions. 
Colormaps of the temperature in longitudinal sections at
$r=\frac{r_o+r_i}{2}$ are depicted. 
The onset of instability occurs 
at $Ra \approx 942.88$ resulting in a spiral flow pattern 
with azimuthal mode number $n=2$. This is stable only in the vicinity of the critical point. 
A small increase in $\Delta T$ leads to a new spiral
state with $n=3$ (figure \ref{Ra_increases} $(a)$), which remains
stable for $1065 \lesssim Ra \lesssim 2485$. Complex spatio-temporal
dynamics emerges as $Ra$ is further increased ($Ra > 2485$). The
sequence is as follows. First, spiral states turn into wavy spiral
flow patterns shown in figures \ref{Ra_increases} $(b)$ and $(c)$,
corresponding to $Ra=2840$ and $Ra=3550$ respectively. Note that the
dominant spiral mode changes back from $n=3$ to $n=2$ as $Ra$ is
increased. These flow patterns are similar to those reported in both
experimental~\cite{LeGoPriMu08} and numerical studies~\cite{ViPo14} 
for large $\Gamma$ and $\eta \approx 0.8$. 
Subsequently increasing $Ra$ above the wavy spiral flow regime 
the flow becomes chaotic (figure \ref{Ra_increases} $(d)$).

The sequence of flow states obtained in the finite case (lower row of
figure~\ref{Ra_increases}) is
qualitatively similar to the periodic case up to the transition to the
convective basic state, which is found to occur at $Ra \approx 4000$, 
in qualitative agreement with criterion~\eqref{eq:crit0.5}. 
Nevertheless, there are quantitative differences. Here the conductive
basic flow (figure~\ref{Ra_increases} $(e)$) remains stable up to
$Ra\approx 1420$, where spiral flow with $n=3$ develops
(figure~\ref{Ra_increases} $(f)$). As the radial heating increases the
dominant spiral mode changes to $n=4$ at $Ra\approx 3550$
(figure~\ref{Ra_increases} $(g)$). 
This increase in $n$ with $Ra$ is driven by the stronger axial flow acting on the fluid because of thermal
buoyancy~\cite{KuBa97}. Further increasing $Ra$ leads to the thinning 
of the thermal boundary layers at the sidewalls and the flow
transitions gradually to the convective laminar flow
(figure~\ref{Ra_increases} $(h)$). 
A similar transition from spiral flow to 
the convective state as $\Delta T$ is increased was experimentally 
observed by~\cite{BaFa89} using an apparatus with $\Gamma=31.5$. 
 We refer the reader to Ref.~\cite{KuBa97} for a detailed description 
of these instabilities.  
       
\begin{figure}
   \begin{center}
     \begin{tabular}{cc}
       $(a)$ & $(b)$\\
       \includegraphics[width=0.48\linewidth]{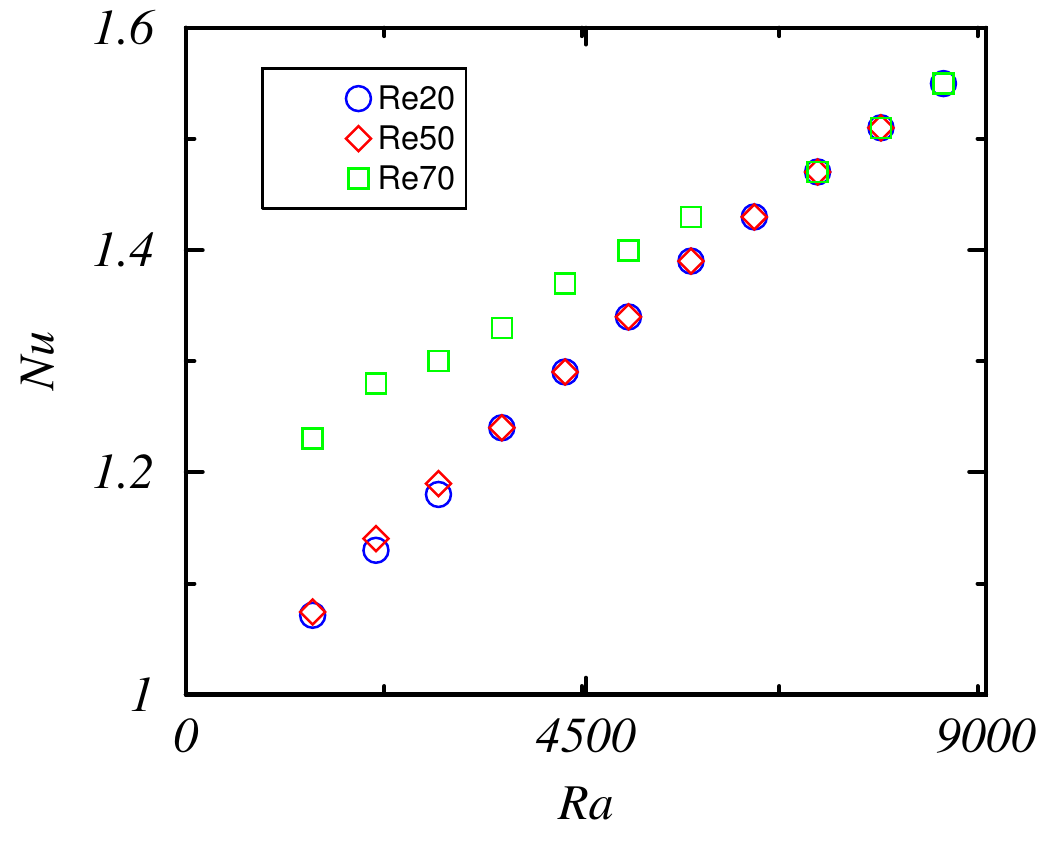}&
       \includegraphics[width=0.48\linewidth]{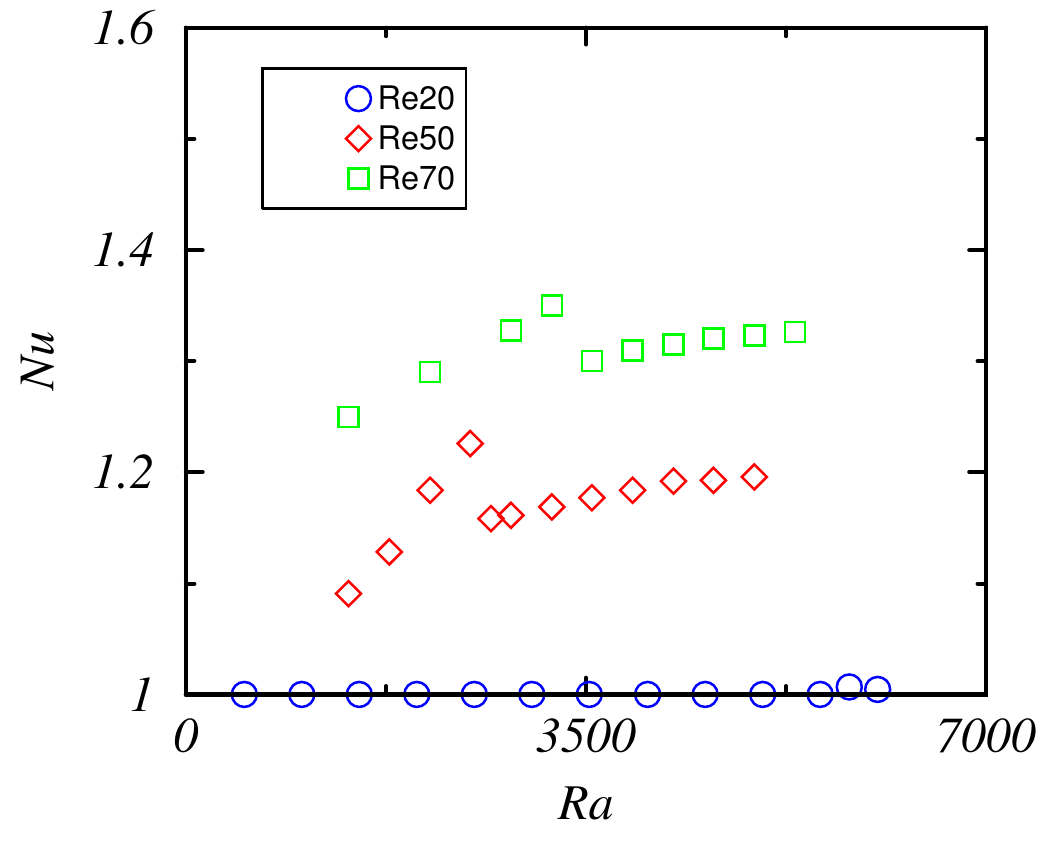}
     \end{tabular}
   \end{center}
   \caption{ Variation of $Nu$ with $Ra$ as a function of $\Rey$ 
     for $\sigma=0.71$ and  $\eta=0.5$ in the
     $(a)$ finite ($\Gamma=10$) and $(b)$ infinite ($L_z=7.48$)  systems.}
   \label{Nuss_Ra}
 \end{figure}

The evolution of the normalised $Nu$ with $Ra$ 
clearly illustrates the different
dynamical behaviours in the finite (figure~\ref{Nuss_Ra} $(a)$) and 
infinite (figure~\ref{Nuss_Ra}  $(b)$) systems. 
The largest differences occur for weak rotation ($\Rey=20$). 
In the infinite case , $Nu=1$ remains constant up to $Ra
\approx 7810$, indicating that heat transport is exclusively
conductive up to reach the onset of instability. 
In contrast,  in the finite case, the Ekman layers
provide an efficient mechanism to transfer heat, 
which results in an almost linear increase of $Nu$ with $Re$. 
Such growth is also observed for spiral flow, $\Rey=50$ and $\Rey=70$. 
When $Ra$ exceeds the value corresponding to the upper
part of the marginal curve in figure~\ref{stability} $(a)$ the flow
becomes again axisymmetric and $Nu$ solely depends on $Ra$. 
As a result, curves corresponding to different values of $\Rey$
collapse. In the infinite case, the behaviour of $Nu$ within the region
of spiral flow, 
$\Rey=50$ and $\Rey=70$, reflects the transitions described above. 
The spiral flow ($Ra \lesssim  2750$) resulting from the primary transition
provides efficient heat transfer, 
leading to a rapid linear growth of $Nu$ as $Ra$ is increased. 
In this flow regime the values of $Nu$ in the infinite case are 
slightly larger than those in the finite case. 
The transition towards wavy spiral flow results in a sudden decrease 
of the convective heat transport, consistently with the results
of~\cite{KeHuCo98}. 
After this initial drop, further increasing $Ra$ towards fully
developed turbulent 
flow is accompanied by a progressive growth in $Nu$, 
which is however significantly lower than the linear growth 
from the spiral flow regime.  

\section{Discussion}\label{sec:discuss}

Heat transfer in laminar flows between two concentric vertical
cylinders can be of conductive or convective nature. 
Regardless of the curvature of the cylinders, parametrised here by the
radius ratio $\eta$, the transition between these regimes is governed
by the Rayleigh number $Ra$ 
and the length-to-gap aspect ratio $\Gamma$. 
In particular, for a given $\Gamma$ there is a critical $Ra$ above
which the laminar state is convective. 
We here determined numerically a linear criterion~\eqref{eq:crit}, 
which applies to all geometries and is independent of the Prandtl
number as long as $\Gamma \gtrsim 15$. 
For smaller $\Gamma$ the transition to the convective state is
 influenced 
by $\sigma$ because of the stronger effect of the Ekman layers.  
Although the criterion was obtained for a fixed Reynolds number
$\Rey=30$, we provided evidence that it appears valid as $Re$ increases. 
This was done by comparing results obtained with axially periodic
boundary conditions, 
for which the laminar flow is always conductive, 
to results obtained with physical (no-slip) boundary conditions.

If the laminar flow is conductive and $Re$ is increased while $Ra$ is
kept fixed, our direct numerical simulations show that the transition 
to chaotic flow is very similar in the periodic and finite cases. 
This was shown by using $\eta=0.5$, $\sigma=0.71$, $Ra=1420$ and
$\Gamma=10$ in the finite case, 
following Kuo and Ball~\cite{KuBa97}.  
All their reported transitions were reproduced here, 
however, we found a notable difference in the onset of spiral flow, 
which occurs for lower $\Rey$ in our simulations and corresponds 
to a critical spiral mode $n=3$ instead of $n=2$. 
This can be explained because Kuo and Ball~\cite{KuBa97} 
neglected the correction of the velocity field in their simulations, 
so their velocity field was not mass conserving. 
Omitting the correction in our simulations we reproduced 
their results and found that its relevance diminishes as $\Rey$
increases, so that the secondary transitions reported 
here coincide with those in~\cite{KuBa97}. 

We further repeated these simulations by using periodic boundary
conditions in a domain of length $L_z=7.48$. 
We found that the onset of instability occurs at slightly lower $\Rey$
than in the finite case, where the Ekman layers act as frictional
layers, slowing down the internal flow and thus delaying the primary 
and secondary instabilities. In contrast, as the $\Rey$ is further
increased the remaining instabilities, culminating in Taylor-vortex
flow, are favoured by the presence of axial end walls. 
In this case the Ekman circulations at the end wall 
enhance the Taylor vortices, as happens in the isothermal case. 

The heat transfer, quantified by the convective Nusselt number
\eqref{eq:Nu}, follows a simple power law ($Nu=0.14\Rey^{0.51}$ and
$Nu=0.13\Rey^{0.53}$ in the finite and infinite cases
respectively). The exponents are in good agreement with the analytical
correlation provided by Dorfman~\cite{dorfman1963}. A similar exponent
($B=0.47$) has also been found by Viazzo and Poncet~\cite{ViPo14} who
numerically explored a region of parameter space similar to 
that in figure~\ref{Nuss_Re}, but using an apparatus
with $\eta=0.8$ and $\Gamma=80$. 
The small discrepancies between the correlations  obtained in the finite and infinite cases 
occurs because of the Ekman layers, whose importance grows as $\Gamma$ is reduced.

If $Re$ is kept fixed while $Ra$ increases, the transition to
chaotic flow and heat transfer of the periodic 
and the finite system remain similar as long as
criterion~\eqref{eq:crit} is approximately satisfied. 
In the finite system, further increases in $Ra$ result in a transition 
to the laminar convective state, whereas the periodic system becomes
gradually more turbulent. 
Thus there are two counter-acting mechanisms as $Ra$ increases. 
Instabilities and turbulence are stimulated because of thermal
driving. However, thermal driving results in a thinning of the thermal
boundary layers because of end-wall effects. 
This quenches turbulence and causes the transition to the laminar
convective state. Noteworthy, at a fixed $Ra$ the laminar convective
state is much more efficient in transferring heat 
than the turbulent flow of the periodic system.

\section{Conclusions}

The role of axial boundaries in laterally heated and rotating flows was investigated with linear stability analyses and direct numerical simulations. Both periodic and no-slip boundary conditions were considered in the axial direction. The main results can be summarised as follows:

\begin{itemize}

\item Criterion~\eqref{eq:crit} determines whether laminar heat transfer in an axially bounded system is of conductive or convective nature. It is independent of Prandtl number and robust with Reynolds number as long as $\Gamma\gtrsim15$.

\item If the laminar state is conductive axially periodic boundary conditions correctly describe the dynamics and heat transfer in laboratory setups. If the laminar state is convective the infinite-cylinder approximation may be used but the basic flow must be modified to include axial stratification~\cite{AlMcF05}. 

\item A variety of flow patterns was observed.  The underlying flow instabilities are driven by the interaction of two physical mechanisms: thermal buoyancy generates an inflectional axial velocity profile, whereas the inner-cylinder rotation promotes centrifugal instability. 

\end{itemize}

\section*{Acknowledgement}
This work was supported by the Spanish Government grant
FIS2013-40880-P and BES-2010-041542.
Part of this work was done during the
visit of J.M. Lopez to the Institute of Science and Research (IST)
in Klosterneuburg (Austria) whose kind hospitality is warmly
appreciated. We thank Red Espa\~nola de
Supercomputaci\'on (RES) for the computational
resources provided.

\section*{References}

\bibliography{local_1}

\end{document}